\documentclass[preprint,twocolumn]{aastex631}

\newcommand{\rsc}[1]{{\color{red}#1}}

\received{2025 September 6}
\revised{revised 2025 November 27}
\accepted{accepted 2025 November 29}

\submitjournal{ApJS}
\usepackage{newtxtext,newtxmath}                    

\usepackage{listings}
\usepackage{graphicx}
\usepackage{float}

\usepackage[T1]{fontenc}

\DeclareRobustCommand{\VAN}[3]{#2}
\let\VANthebibliography\thebibliography
\def\thebibliography{\DeclareRobustCommand{\VAN}[3]{##3}\VANthebibliography}

\usepackage{amsmath}	
\usepackage{hyperref}

\newcommand{\MS}{\ifmmode{\,}\else\thinspace\fi{\rm M}\ifmmode_{\odot}\else$_{\odot}$\fi}
\newcommand{\LS}{\ifmmode{\,}\else\thinspace\fi{\rm L}\ifmmode_{\odot}\else$_{\odot}$\fi}
\newcommand{\RS}{\ifmmode{\,}\else\thinspace\fi{\rm R}\ifmmode_{\odot}\else$_{\odot}$\fi}
\newcommand{\npg}{\ifmmode{^{14}\rm N (\rm p,\gamma) ^{15} \rm O}\else{$^{14}\rm N (\rm p,\gamma)^{15}\rm O$}\fi}
\newcommand{\cag}{\ifmmode{^{12}\rm C (\alpha,\gamma) ^{16} \rm O}\else{$^{12}\rm C (\alpha,\gamma)^{16}\rm O$}\fi}
\newcommand{\cc}{\ifmmode{^{12}\rm C}\else{$^{12}\rm C$}\fi} 
\newcommand{\nn}{\ifmmode{^{14}\rm N}\else{$^{14}\rm N$}\fi} 
\newcommand{\oo}{\ifmmode{^{16}\rm O}\else{$^{16}\rm O$ }\fi} 

\newcommand{\ooo}{\ifmmode{^{15}\rm O}\else{$^{15}\rm O$}\fi} 
\defcitealias{Ziolkowska-2024}{Paper~I}

\begin{document}

\title[Toward a Comprehensive Grid of Cepheid Models]{Toward a Comprehensive Grid of Cepheid Models with \texttt{MESA}\\ II. Impact of Physical and Numerical Assumptions on Elemental Abundances}

\correspondingauthor{Oliwia Zi\'{o}\l{}kowska}
\email{oliwiakz@camk.edu.pl}
\author{O. Zi\'{o}\l{}kowska}
\affiliation{Nicolaus Copernicus Astronomical Centre, Polish Academy of Sciences, Bartycka 18, 00-716 Warszawa, Poland}

\author{R. Smolec}
\affiliation{Nicolaus Copernicus Astronomical Centre, Polish Academy of Sciences, Bartycka 18, 00-716 Warszawa, Poland}

\author{A. Thoul}
\affiliation{Space sciences, Technologies and Astrophysics Research (STAR) Institute, Université de Li\`{e}ge, All\'{e}e du 6 Ao\^{o}t 19C, Bat. B5C, B-4000 Li\`{e}ge, Belgium}

\author{R. Singh Rathour}
\affiliation{Nicolaus Copernicus Astronomical Centre, Polish Academy of Sciences, Bartycka 18, 00-716 Warszawa, Poland}

\author{V. Hocd\'{e}}
\affiliation{Nicolaus Copernicus Astronomical Centre, Polish Academy of Sciences, Bartycka 18, 00-716 Warszawa, Poland}

\begin{abstract}

Modern tools for modeling stellar evolution, such as \texttt{MESA} (Modules for Experiments in Stellar Astrophysics), offer state-of-the-art implementations of stellar theories. However, this parametric approach introduces many free parameters that are often not constrained by observations. This is particularly important for evolved stars, like classical Cepheids, because uncertainties increase with evolution time. In previous work, we studied the effect of varying microphysics, including solar abundance mixtures, nuclear networks, atmosphere models, mixing-length prescriptions, treatments of convective boundaries, and numerical setup on evolutionary tracks. Here, we extend this analysis to the surface abundances of the dominant elements H, He, C, N, O, Ne, and Mg.
We establish a reference model and 22 variants for each mass and metallicity, $M$/$Z$, evolving them from the Zero-Age Main Sequence to central helium exhaustion. Masses between 2–8\MS\ and metallicities $Z=0.0014, 0.004, 0.014$ are explored, spanning the range of classical Cepheids. Both canonical and overshooting models are computed and compared.
We find that uncertainties in surface abundances are generally small, arising mainly from variations in the depth of the convective envelope during the first dredge-up. The size of the convective envelope is sensitive to many aspects, including mass and metallicity. The central C/O ratio, relevant for white dwarf evolution, can vary by $\sim$0.15, driven largely by convective boundary treatments or by modifying the \cag\ reaction rate during helium burning. Surface and central abundances for the considered models at several benchmark points during the evolution are provided online.

\end{abstract}

\keywords{Cepheid variable stars (218)--- Stellar evolutionary tracks (1600)--- Chemical abundances (224)}

\section{Introduction}

Stellar evolution codes are a crucial and widely used tool for contemporary astrophysics, allowing us to model stars across the whole range of masses and chemical compositions, including phenomena such as rotation, mass transfer, pulsations, and others. Comparing the results of the models with observational inferences improves our understanding of the inner workings of stars and in turn, helps us to refine our modelling tools. However, the most commonly used evolution codes are one-dimensional and rely on simplified physical assumptions, e.g., describing the convective transport of energy via mixing-length theory \citep[MLT;][]{Bohm-Vitense1958}. Moreover, there are gaps in our understanding of aspects of stellar evolution, such as the relative distribution of heavy elements in the Sun \citep[see e.g.,][]{Magg-2022}, or nuclear reaction rates \citep[see e.g.,][]{Morel-2010}. Additionally, every stellar evolution code is characterized by a specific numerical setup, with parameters controlling the precision of solvers, temporal and spatial resolution, and more. All these aspects, different codes, their physical and numerical configuration, contribute to the rise of modelling uncertainties \citep[e.g.,][]{Ziolkowska-2024, Li-2025}. The latter two issues, namely the physical and numerical configuration within a chosen evolution code, were addressed in our first paper \citep[][hereafter Paper~I]{Ziolkowska-2024}, in which we estimated the systematic uncertainties of luminosity, effective temperature, and age in \texttt{MESA} models of intermediate-mass stars. We had shown that the late-type stars are affected more than those on the main sequence. A particularly sensitive part of the evolutionary track is the blue loop, where Cepheid pulsation phase can occur. Here, we continue our work by analyzing how the models are affected in terms of the chemical abundances.

The recent efforts of increasing precision and accuracy of spectroscopic abundances of FGK-type stars were provided in \citet{Jofre-2019}. Many limitations affect spectral abundances determination, starting from observational issues, like signal-to-noise ratio and resolution, physical assumptions on the line formation ( assuming local thermodynamic equilibrium or not, one-dimensional or three-dimensional geometry), and their transition probabilities, asymmetries due to convection, non-radial pulsations, spots, or blends, and the analysis method, e.g., normalization to the continuum.

Thanks to the availability of reliable parallax data from Gaia DR2 \citep{GaiaDR2-2018} and DR3 \citep{GaiaDR3-2023}, many studies have appeared on the abundance gradients of Cepheids in the Milky Way \citep{daSilva-2023, Lemasle-2022, Skowron-2019, Luck-2018}. In one of these studies, \citet{Trentin-2024}, as a part of their C-metaLL survey, provided a sample of abundances for 180 Galactic Cepheids, covering a broad range of metallicities and enlarging their previous sample to 292 Cepheids. 

Several works have discussed chemical abundances from stellar evolution models, mostly in the context of stellar rotation and associated mixing, usually focused on massive stars \citep[e.g.,][]{Heger-2000}. 
The influence of rotation rates on Cepheid abundances has been analyzed, e.g., by \citet{Anderson-2014}, where they argue that enhanced surface abundances are not a reliable tool to distinguish between first-crossing and blue-loop Cepheids. Regarding specific elements, recent papers, discussing $^7$Li, [C/N], $^{12}$C/$^{13}$C, that are sensitive to mixing on the RGB, specifically near the RGB bump, due to thermohaline mixing, include, e.g. \citet{Tayar-2022}, who provide a concise summary of the literature on the subject and use MESA to explore explanations beyond the usual thermohaline mixing explanation. Similarly, \citet{Schwab-2020} provides a new explanation for extra mixing induced by helium sub-flash. MESA models were also used by \citet{Fraser-2022}, together with observational data to construct a framework to compare different model thermohaline mixing prescription. \citet{Tautvaisiene-2016} compare their spectroscopically determined CNO abundances and carbon isotope ratio of evolved stars with PARSEC isochrones. \citet{Tsiatsiou-2025} investigated fluorine abundance in star across a wide range of mass and metallicity in stellar evolution models, including rotation. In a series of papers \citet{Charbonnel-2010, Legarde-2011, Legarde-2012} authors discuss abundances of various elements of low- and intermediate-mass models up to AGB, including thermohaline and rotational mixing.

In pre-AGB evolution, the most notable change of photospheric elemental abundances occurs during the first dredge-up on the red giant branch, where the convective envelope reaches the matter partly processed by the CNO cycle and brings it to the surface, resulting in carbon deficit and nitrogen enhancement as compared to the Sun \citep[e.g.,][]{Iben-1967, Luck-1978, Luck-2018}. In the case of our reference models, the changes in surface abundances due to the first dredge-up are on average $\Delta$[N/H]$=$0.33~dex, $\Delta$[C/H]$=-$0.16~dex, $\Delta$[He/H]$=\Delta$[O/H]=0.01~dex, and $\Delta$[Ne/H]$=\Delta$[Mg/H]$=$0.003~dex (average over mass and metallicity of the considered models). Our work focuses on the impact of varying physical and numerical assumptions of our code of choice on the surface abundances and how they change during the evolution, in particular how changes due to the first dredge up are affected. We omit rotation and mass loss, which will be studied separately. We use a publicly available, 1D, open source stellar evolution code, Modules for Experiments in Stellar Astrophysics, \texttt{MESA} \citep{Paxton-2011, Paxton-2013, Paxton-2015, Paxton-2018, Paxton-2019, Jermyn-2023}. We first establish a reference set of models with which we compare our modified models. In those, we vary scaled solar metal mixture, nuclear network, atmosphere model, MLT formalism, convective boundary determination criteria, temporal and spatial resolution, and the method of interpolation of the opacity tables. We compare the abundances on eight evolutionary benchmark points on the tracks. We focus on stable isotopes in our nuclear network, which contains He, C, N, O, Ne and Mg. We discuss in detail the cases where abundance predictions differ from the reference ones by at least 0.01~dex, which is the typical precision given in the literature \citep[e.g.,][]{Asplund-2009}. We also discuss central C/O ratio at the end of core helium burning, which is important and has consequences on later evolutionary stages, such as white dwarf evolution \citep[e.g.,][]{Umeda-1999}.

This paper is the second in a series in which we address several problems regarding evolutionary modeling of classical Cepheids. In \citetalias[]{Ziolkowska-2024} we provided foundations for evolutionary calculations of Cepheid tracks. In the forthcoming work, we will study how mass loss, rotation, and overshooting from the envelope, as well as from the core, affect evolutionary tracks of medium-mass stars.

The structure of the paper is the following. In Sect.~\ref{sec:methods} we detail our methodology. In Sect.~\ref{sec:results} we discuss the models that predict noticeable differences in the abundances. Section~\ref{sec:discussion} presents the discussion and summary. We provide an exemplary reference model inlist in the Appendix~\ref{appendix:inlist}. Tables containing the abundances for all the isotopes in our nuclear network are available online. We provide \texttt{MESA} history files containing all the surface and central abundances from our model grid. In the provided tracks, many quantities are included, in particular central and surface mass fractions of all the traced isotopes in our nuclear network, basic properties, like age, luminosity, effective temperature, radius, surface gravity, and several central parameters: pressure, temperature, molecular weight.

\section{Methods}
In this section, we introduce our methodology and evolutionary assumptions, describe how we quantify abundances, and discuss issues encountered during the modeling.
\label{sec:methods}
Here we introduce our methodology, evolutionary assumptions, and explain how we auantify abundances, discuss some issues that occured during modeling. 

\subsection{Evolutionary Calculations}

\begin{figure*}
    \centering
    \includegraphics[width=.8\linewidth]{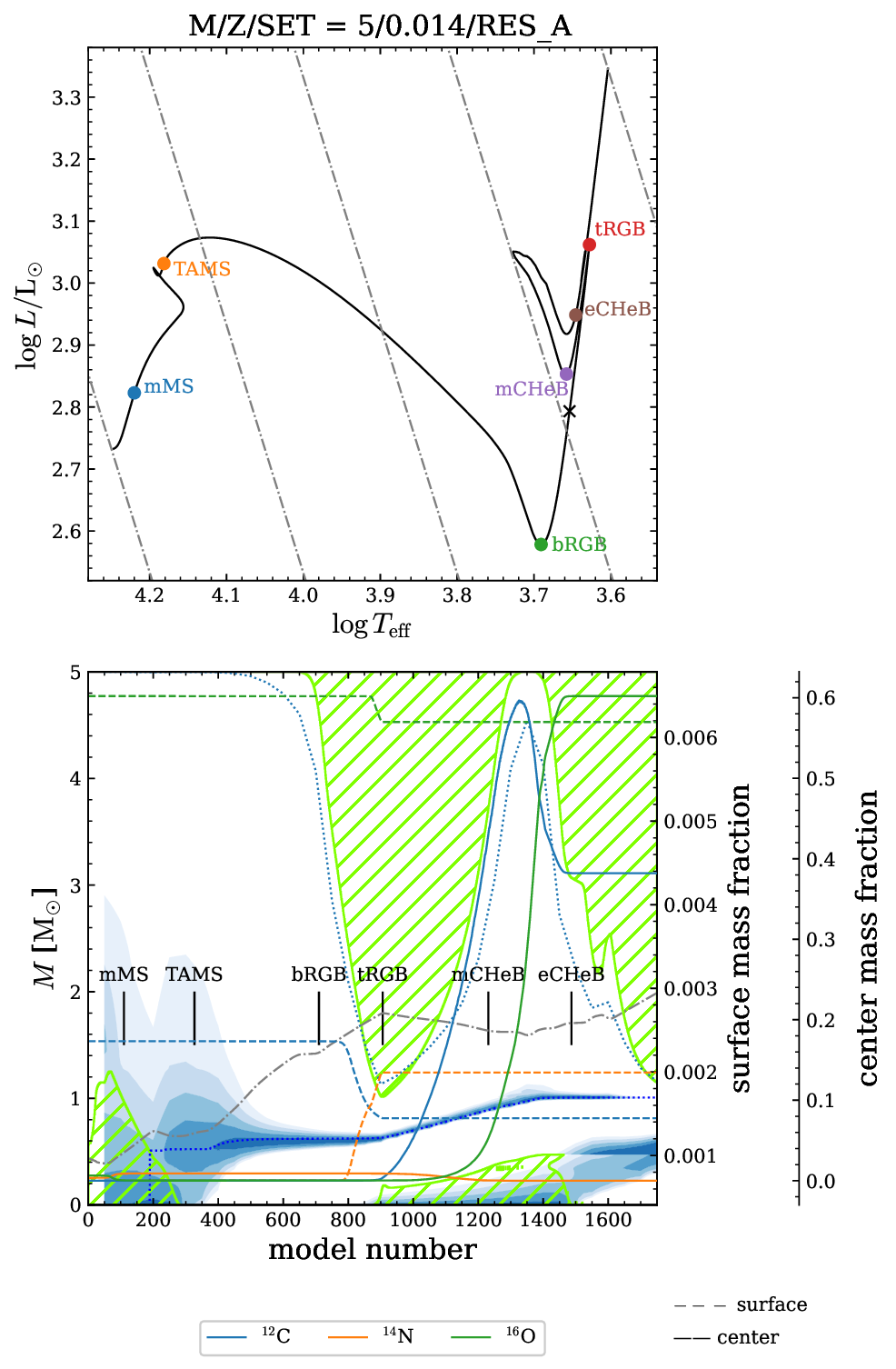}
    \caption{Evolution of a 5\MS, $Z=0.014$ reference model on the HR diagram (left panel) and a corresponding Kippenhahn diagram (right panel). 
    Left: Evolutionary track is shown with color-coded benchmark points. The black cross on the RGB marks the model investigated further with profiles in Fig.~\ref{fig:Tmix-profiles}. With dash-dotted lines, we show the isoradial lines with $\log R$/\RS=0.4, 0.8, 1.2, 1.6, 2.0.
    Right: The corresponding Kippenhahn diagram, showing the interior structure of the model during evolution (time expressed in model number). The green hatched regions mark convection, and the blue-shaded ones mark regions of efficient nuclear burning. We indicate the location of the evolutionary benchmark points (and the model marked with a cross on the HR diagram) with vertical lines and labels. The radius, $\log R / \RS$, vs. model number is shown with a dash-dotted line with the same scale as the mass coordinate on the left axis. Surface (dashed lines) and central (solid lines) mass fractions of $^{12}$C, $^{14}$N, and $^{16}$O are shown with scales given on the right side of the figure. The dotted line marks temperature equal to $10^6$\,K.}
    \label{fig:kip0}
\end{figure*}

The evolutionary calculations presented in this study are constructed using the same physical setup as in \citetalias{Ziolkowska-2024} with one exception, concerning the treatment of surface layers, which demand a more detailed treatment, as discussed in Sect.~\ref{sec:rad}. We use \texttt{MESA}, version r-21.12.1. Our models span masses from $2$ to $8\MS$, with a step of $1\MS$, with three different metallicities corresponding roughly to $\rm [Fe/H]=0.0, -0.5, -1.0$. Evolutionary phases, from Zero-Age Main Sequence (ZAMS) till the end of core-helium burning (CHeB), are calculated. In particular, we distinguish eight evolutionary benchmark points at which we can compare models with given mass and metallicity, $M/Z$. The points are the middle of the main sequence (mMS), Terminal-Age Main-Sequence (TAMS), base and tip of the red giant branch (bRGB, tRGB), beginning, middle, and end of core Helium burning (bCHeB, mCHeB, eCHeB), and middle of the Instability Strip (mIS). In \citetalias[tab.~3,]{Ziolkowska-2024} we provide the exact definitions of those points. Regarding the mIS point, the IS was calculated for six cases -- three metallicity values, with, and without overshoot. It was based on the reference tracks, with masses from $2$ to $8\MS$, with a step of $0.5\MS$. We used Radial Stellar Pulsations (RSP) module of \texttt{MESA}, using the simplest convective envelope parameters (set A) to compute linear growth rate of the fundamental mode ($\gamma_{\rm F}$) along each track and then interpolated to where $\gamma_{\rm F}=0$. We stress however, that the location of the IS and hence the precise location of the mIS point does not affect our results at all. mIS corresponds to a slow, core-helium burning stage {\rsc{of}} evolution during which no changes of surface abundances occur. A representative evolutionary track and a corresponding Kippenhahn diagram, both with marked benchmark points (except mIS, as the model does not reach the center of IS) are presented in Fig.~\ref{fig:kip0}. Mass loss and rotation are neglected in this study. The effects of convective core overshooting on the main sequence are discussed in Sect.~\ref{ssec:overshoot}. 

In this study, we neglect atomic diffusion as it does not affect our evolutionary tracks, see \citetalias{Ziolkowska-2024}. In \citet{Alecian-2023}, the authors present main sequence abundance profiles for a 5\MS\ model with diffusion. Abundances are affected slightly, but not as much as in the case of lower mass stars. Inclusion of atomic diffusion significantly affects the time of computation, and we choose to neglect it, similarly to previous works on intermediate-mass stars \citep[e.g.,][]{Choi-2016}.

\subsection{Reference Model -- Adopted Physics}

The setup of the reference model is detailed in \citetalias{Ziolkowska-2024}. Here we present a short description of the reference model physics. We adopt three initial metallicity ($Z$) values of $0.014$, $0.004$ and $0.0014$. We calculate $X,Y$ and $\rm [Fe/H]$ based on $Z$ using primordial helium abundance $Y_{\rm p} = 0.2485$ from \cite{Komatsu-2011}, and helium enrichment $\Delta Y/\Delta Z=1.5$. The solar-scaled mixture of elements comes from \citet{Asplund-2009}. Opacity comes from cubic interpolation of different tables: OPAL \citep{Iglesias-1993, Iglesias-1996} and \citet{Ferguson-2005} tables for lower temperatures. Type 2 tables, which include enhanced C and O abundances, are used during and after core-helium burning. The boundary atmosphere conditions are set by PHOENIX tables from \citet{Hauschildt-1999a, Hauschildt-1999b} and \citet{Castelli-2003} models. The MLT formalism comes from \citet{Henyey1965}. Boundaries of convective regions are determined by the Schwarzschild criterion and predictive mixing scheme (PM) in the core \citep{Paxton-2018}, while in the envelope we use the classic sign change algorithm. The mixing length parameter, $\alpha_{\rm MLT}$=1.77, was calibrated to fit the solar age and photospheric abundances (see Sect.~2.3 of \citetalias[]{Ziolkowska-2024}). The equation of state tables used are OPAL \citep{Rogers-2002}, SCVH \citep{Saumon-1995}, HELM \citep{Timmes-2000}, PC \citep{Pothekin-2010} and Skye \citep{Jermyn-2021}. Detailed parameter space where each of these EOS is used can be found in \texttt{MESA}'s documentation\footnote{\url{https://docs.mesastar.org/en/release-r21.12.1/eos/overview.html}}. 

We present the complete inlist (input parameter file for \texttt{MESA}) of the reference model in Appendix~\ref{appendix:inlist}.

\subsection{Initial Setup for Isotopes}

In the reference model, we use a nuclear reaction net {\texttt{pp\_and\_cno\_extras.net}} that explicitly tracks 25 isotopes ($^1$H, $^2$H, $^3$He, $^4$He, $^{7}$Li, $^{7}$Be, $^{8}$B, $^{12}$C, $^{13}$C, $^{13}$N, $^{14}$N, $^{15}$N, $^{14}$O, $^{15}$O, $^{16}$O, $^{17}$O, $^{18}$O, $^{17}$F, $^{18}$F, $^{19}$F, $^{18}$Ne, $^{19}$Ne, $^{20}$Ne, $^{22}$Mg, $^{24}$Mg) and includes reactions of the proton-proton (p-p) chain and CNO cycle as well as helium burning.

Before the evolutionary calculations begin, the mass fraction of each element is set according to the chosen reference mixture of elements (A09, GS98 or GN93, corresponding to \citet{Asplund-2009}, \citet{GS-98}, \citet{GN-93}). Then, for each element, the mass fraction is split among specific isotopes based on the solar system isotopic ratios provided by \citet[][their tab.~6]{Lodders-2003}.

\begin{figure}
    \centering
    \includegraphics[width=.9\linewidth]{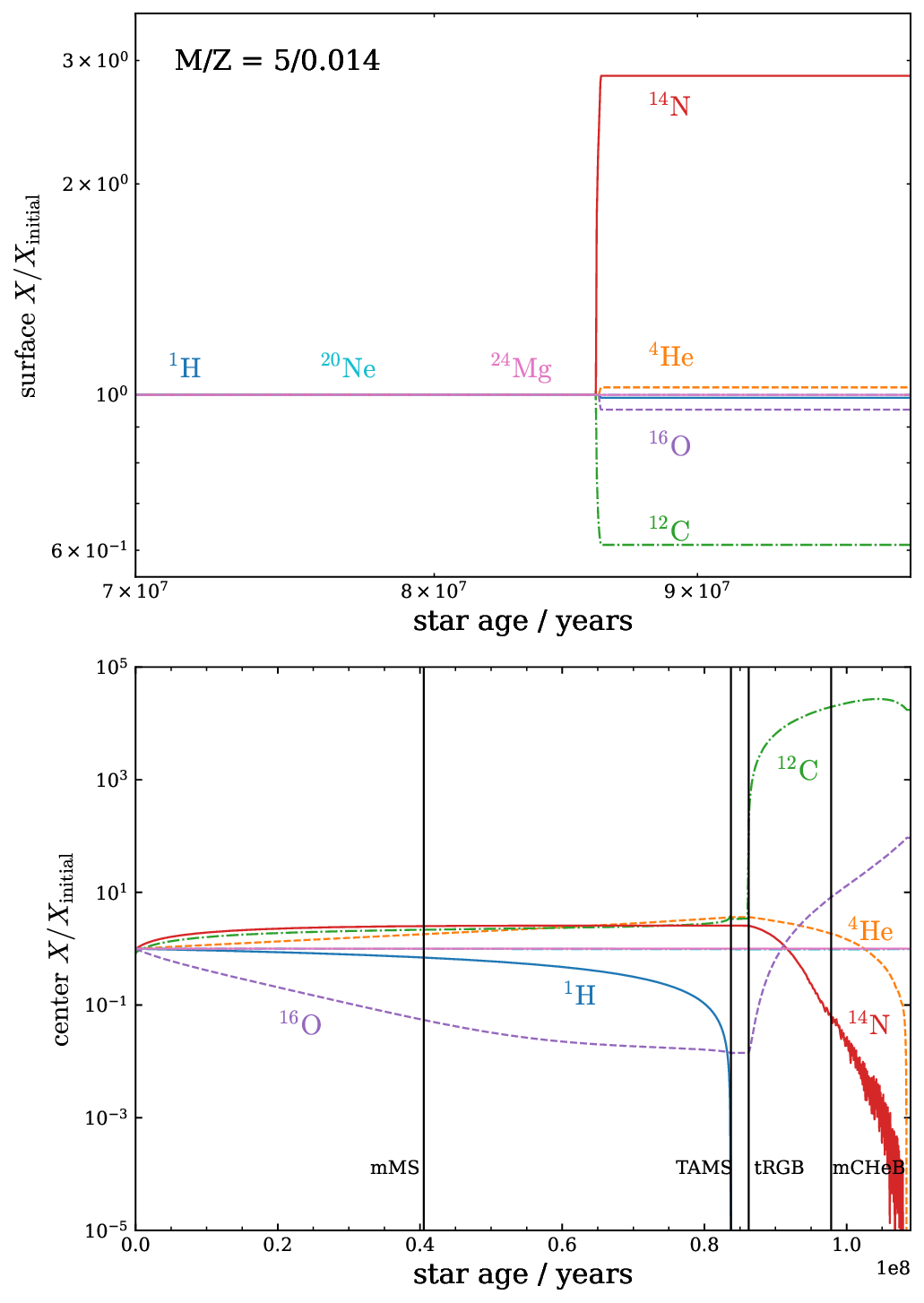}
    \caption{Surface (top) and central (bottom) mass fractions relative to their initial values as a function of time for the $5\MS$, $Z=0.014$ canonical model. The top panel shows a zoom near the first dredge-up, while the bottom panel shows the whole evolution until AGB.}
    \label{fig:massfracs}
\end{figure}

Our calculations start at ZAMS and not at pre-main sequence. This has very few consequences for the tracks on the HR diagram; however, it has implications for the initial ZAMS composition of the models. To build a ZAMS model, \texttt{MESA} uses a sequence of pre-computed ZAMS models with $Z=0.02$ and then gradually changes (or ‘relaxes’) $Z$ and $Y$ to the desired values. These pre-computed models were constructed using a limited nuclear network, \texttt{basic.net}, containing the most abundant elements. When the net is extended, abundances of isotopes of the elements included in \texttt{basic.net}, e.g., of $^3$He or $^{13}$C, are set according to \cite{Lodders-2003}. Most of the other light elements are unstable; they are created and maintain equilibrium abundance only when the reaction chains containing them are ongoing. Lithium is different. Most of it, in the central regions of a star, is burned during pre-main sequence evolution, and its influence on the nuclear evolution of the star is negligible (still, it takes part e.g., in pp cycle). However, at the surface, the abundance of lithium on the ZAMS can be significant and subject to further evolution, for example through the dredge-up process. Studies for which the surface abundance of lithium or other light elements is important should take full account of the pre-main sequence phase. Consequently, our analysis is focused on these most abundant elements, which are $^1$H, $^4$He, $^{12}$C, $^{14}$N, $^{16}$O, $^{20}$Ne, $^{24}$Mg. We save the mass fraction values for these isotopes on
the surface of the star and in the center through the entire
evolution with \texttt{MESA} history files and additionally
their distribution within the star at a given time with \texttt{MESA}
profile files.  We show their evolution on the surface and in the center for a $5\MS, Z=0.014$ reference model in Fig.~\ref{fig:massfracs}.

\subsection{Thin Radiative Surface Layer Issue}
\label{sec:rad}

In principle, the convection on the RGB should be reaching all the way to the surface. In some of our models, however, we observe surface radiative layers, covering 30--60 surface zones, which is equivalent to $\sim$0.009\RS, or $\sim$$10^{-6}$\MS\ in mass coordinate, all within the optical depth $\tau\sim1$, which roughly corresponds to the photosphere of a star.

\begin{figure*}
    \centering
    \includegraphics[width=\hsize]{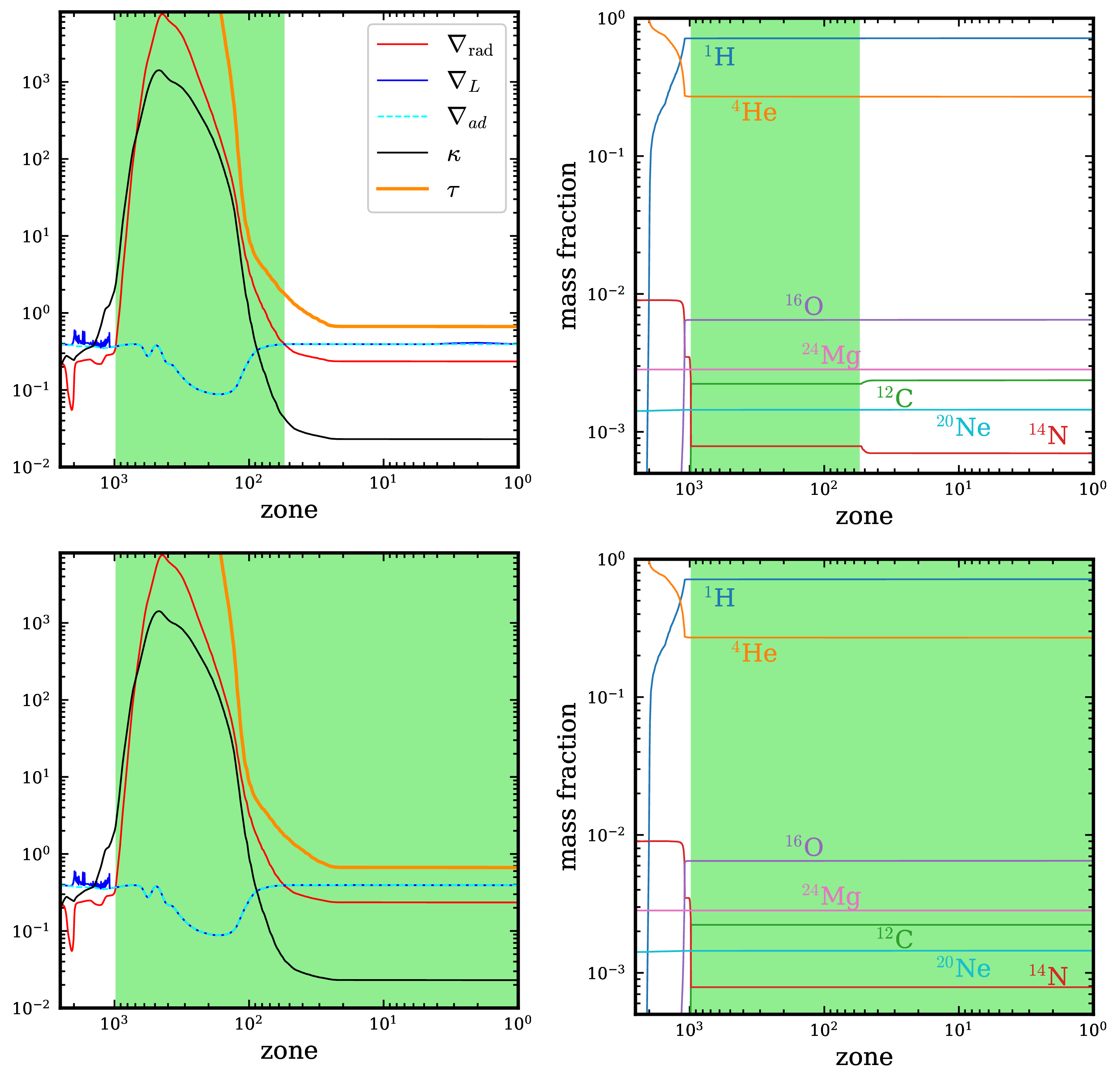}
    \caption{Profiles of the 5\MS, $Z=0.014$ model marked on the HR diagram (Fig.~\ref{fig:kip0}) with a cross. Left: temperature gradients (radiative, Ledoux, adiabatic), opacity and optical depth, as a function of zone (starting at one from the surface and increasing towards the center). Right: Mass fractions of the analysed isotopes as a function of zone number. 
    The top panels show the profiles before removing the surface radiative shell, and the lower ones -- after.
    Green area marks the convective envelope.}
    \label{fig:Tmix-profiles}
\end{figure*}

In the upper panels of Fig.~\ref{fig:Tmix-profiles}, we show profiles of physical quantities (left panels) and abundances (right panels) for the $5\MS/Z=0.014$ model on the RGB (model location is marked with a cross on Fig.~\ref{fig:kip0}). The x-axis is the zone number, starting with 1 at the surface, and increasing towards the center. The convective (green-shaded) region does not reach the 50 surface zones, as close to the surface, the radiative temperature gradient drops below the adiabatic one; a consequence of strong opacity decrease towards the surface. Thus, the surface abundances may be slightly affected, depending on whether the radiative surface layer developed while the convective envelope was still increasing in depth, mixing the regions that underwent nuclear processing. This is exactly the case for the presented model (observe its location on the Kippenhahn diagram in Fig.~\ref{fig:kip0}).

As a consequence, a composition of a very thin (optical depth below 1) surface layer becomes frozen, while directly underneath, it still evolves. This is well visible in top right panel of Fig.~\ref{fig:Tmix-profiles} for $^{12}$C and $^{14}$N. In a real star, a slight amount of any mixing process (e.g., due to overshooting or rotation) would make the surface layers well-mixed. The issue we discuss here is hence a consequence of the adopted physical assumptions. We decided to mix this thin radiative region artificially, with a control builtin in \texttt{MESA} for that purpose, \texttt{T\_mix\_limit}. The effect is shown in the bottom panels of Fig.~\ref{fig:Tmix-profiles} where we used {\texttt{T\_mix\_limit=1d6}}. This checks if there are any convective regions below the specified temperature value ($10^6$\, K in this case) and extends these regions to the very surface of the star. We adopt the value of $10^6$\, K (see blue dotted line in the Kippenhahn diagram in Fig.~\ref{fig:kip0}) after checking that the tracks remain unchanged with this value for all considered $M/Z$. The same control was used by \citet{Schwab-2020} for studying lithium abundance in red clump stars.

\subsection{Modified Models}

The physical and numerical setup of the modified models is presented in Tab.~\ref{tab:labels}. In our sets of models we modify interpolation scheme of opacity tables (the set of models with different settings is referred to as INT in the following), scaled solar mixtures of heavy elements (MIX), nuclear networks (NET), atmosphere models (ATM), mixing length theory formalisms (MLT), convective boundary determination criteria (CONV), spatial and temporal resolution (RES). Detailed settings of models comprising a given set are given in Tab.~\ref{tab:labels}.
In the case of MIX, the opacity tables used were constructed using the associated solar compositions.

Concerning overshooting, we consider two cases: canonical models (with no overshooting), and models with moderate overshooting from a hydrogen-burning convective core on the main sequence. We use the exponential prescription by \citet{Herwig-2000} \citep[for \texttt{MESA} implementation see][]{Paxton-2011} with $f=0.02$. All the models mentioned in Tab.~\ref{tab:labels} were calculated with and without overshooting, and its effect is discussed in Sect.~\ref{ssec:overshoot}. 

\begin{table*}
\caption{Characteristics of various sets of models considered in this paper: INT, MIX, NET, ATM, MLT, CONV, and RES. The reference model is highlighted with boldface font and is always labeled with `A'. In a given set, its relevant controls are only varied (e.g., controls related to atmospheric boundary conditions in ATM), while all other controls are set to their reference values.}
\begin{tabular}{ll}
\hline
Set & Varied options \\ \hline 
{\textbf{INT\_A}} & cubic $X/Z$ interpolation of opacity tables \\ 
INT\_B & linear $X/Z$ interpolation of opacity tables\\ \hline
{\textbf{MIX\_A}} & scaled solar mixture based on \citet{Asplund-2009} \\
MIX\_B & scaled solar mixture based on \citet{GS-98} \\
MIX\_C & scaled solar mixture based on \citet{GN-93} \\ \hline
{\textbf{NET\_A}} & \cag\ from \citet{Kunz-2002} +  \npg\ from \citet{Cyburt-2010} +  pp\_and\_cno\_extras.net\\
NET\_B & \cag\ from \citet{Kunz-2002} +   \npg\ from \citet{Cyburt-2010} +  mesa49.net \\
NET\_C & \cag\ from \citet{Angulo-1999} +  \npg\ from \citet{Cyburt-2010} +  pp\_and\_cno\_extras.net\\
NET\_D & \cag\ from \citet{Kunz-2002} +  \npg\ from \citet{Angulo-1999} +  pp\_and\_cno\_extras.net\\
NET\_E & \cag\ from \citet{Angulo-1999} +  \npg\ from \citet{Angulo-1999} +  pp\_and\_cno\_extras.net\\\hline 
{\textbf{ATM\_A}} & model atmosphere tables  \citep{Hauschildt-1999a,Hauschildt-1999b, Castelli-2003} \\
ATM\_B & $T$-$\tau$ relation Eddington \\
ATM\_C & $T$-$\tau$ relation Krishna\_Swamy \citep{Krishna-Swamy-1966}\\
ATM\_D & $T$-$\tau$ relation solar\_Hopf \citep{Paxton-2013}\\ 
ATM\_E & $T$-$\tau$ relation Trampedach\_solar \citep{Ball-2021, Trampedach-2014} \\ \hline
{\textbf{MLT\_A}} & Henyey \citep{Henyey1965} \\
MLT\_B & ML1 \citep{Bohm-Vitense1958} \\
MLT\_C & Cox \citep{Cox-1968} \\
MLT\_D & Mihalas \citep{Mihalas-1978} \\ \hline
{\textbf{CONV\_A}} & predictive mixing + Schwarzschild criterion \\
CONV\_B & predictive mixing + Ledoux criterion\\
CONV\_C & sign change algorithm + Schwarzschild criterion \\
CONV\_D & predictive mixing + Schwarzschild criterion + including predictive mixing in the envelope\\ \hline
{\textbf{RES\_A}} & \texttt{time\_delta\_coeff = 0.50} + \texttt{mesh\_delta\_coeff = 0.50} \\ 
RES\_B & \texttt{time\_delta\_coeff = 0.25} + \texttt{mesh\_delta\_coeff = 0.50} \\
RES\_C & \texttt{time\_delta\_coeff = 0.50} + \texttt{mesh\_delta\_coeff = 0.25} \\
RES\_D & \texttt{time\_delta\_coeff = 0.25} + \texttt{mesh\_delta\_coeff = 0.25} \\
RES\_E & \texttt{time\_delta\_coeff = 1.00} + \texttt{mesh\_delta\_coeff = 1.00} + default \texttt{MESA} resolution controls\\ \hline
\end{tabular}
\label{tab:labels}
\end{table*}

\subsection{Quantifying Abundances}

We define the surface abundance of an element X as $\log (\rm X/\rm H) + 12$, where X is the mass fraction of the outermost zone in the model and H is the mass fraction of hydrogen of the same zone.

For further analysis, for each of the remaining elements, He, C, N, O, Ne, Mg, we focus on the most abundant isotope. We save the abundances at the eight evolutionary benchmark points that are marked in Fig~\ref{fig:kip0}. We look into the models with $>0.01$~dex abundance difference between the reference and modified model -- investigating their profiles, where the abundance difference is defined as $$\rm \Delta[X/H]=[X/H]_{ reference}-[X/H]_{ modified}\,.$$

\section{Results}
\label{sec:results}

Our results are available through Zenodo under an open-source Creative Commons Attribution license at doi: \dataset[https://zenodo.org/uploads/15363589]{https://zenodo.org/uploads/15363589}. We provide tables of elemental abundances of our models at eight evolutionary benchmark points: mMS, TAMS, bRGB, tRGB, bCHeB, mCHeB, eCHeB, and mIS, as illustrated with Tab.~\ref{tab:abund} for eCHeB. For each combination of mass and metallicity, 22 rows are present, each corresponding to a modified model, according to the labels in Tab.~\ref{tab:labels}. Columns contain mass, metallicity, model label, central C/O ratio, surface mass fraction of hydrogen, and abundances of He, C, N, O, Ne, and Mg. A similar collection of tables for central mass fractions of the same isotopes is present online. Finally, both surface and central abundance tables are included in two variants, without overshooting and with moderate overshooting from the convective core on the MS. 

The evolutionary tracks are available as \texttt{MESA} history files containing multiple columns with central and surface mass fractions of all the isotopes traced in our nuclear network, basic properties, like age, luminosity, effective temperature, radius, surface gravity, and several central parameters: pressure, temperature, and molecular weight. 

The main focus of this work is the surface abundances and the central $^{12}$C/$^{16}$O ratio. We compare the surface abundances between the reference and modified models for every $M/Z$ combination. For each model in which surface abundance of one of the considered elements differs from the value in the reference model by $\geq$0.01~dex, abundance profiles and internal structure are inspected.

The evolution of surface and central abundances is shown in Fig.~\ref{fig:massfracs}, in the top and bottom panels, respectively, for a representative, 5\MS, $Z=0.014$ model. At the initial evolutionary stages, the surface abundance differences between the reference and the modified models are negligible. The envelope is radiative and no mixing occurs; ZAMS composition is preserved. The step-like change in surface abundances corresponds to the first dredge-up during the red giant branch phase; for the models of 2\MS\ it occurs on the RGB, followed by a luminosity bump, and for the higher masses, exactly at tRGB. The elements most affected by the first dredge-up are N and C, followed by O, He and H. Ne and Mg do not change noticeably (see top panel in Fig.~\ref{fig:massfracs}). In the case of the central abundances (bottom panel in Fig.~\ref{fig:massfracs}, with mMS, TAMS, tRGB, and mCHeB points denoted by vertical lines), the highest changes in C, N, and O occur also after the tRGB, due to the onset of helium burning reactions in the emerging convective core.

In the next sections, we focus on the surface abundances. The maximal depth of the convective envelope, which determines the amount of mixing during the first dredge up, is affected by many factors, constituting the main source of the modeled surface abundance uncertainty.

\begin{table*}[]
\caption{A section of an online table containing surface abundances at the end of core helium burning (eCHeB). Columns contain mass, metallicity, designation of a model set, the central C/O ratio, surface hydrogen mass fraction, abundances of He, C, N, O, Ne, and Mg.}
\begin{tabular}{lllllllllll} \hline \hline
mass & Z & set & $^{12}$C$/^{16}$O$_{\rm center}$ & $^1$H$_{\rm surface}$ & [He/H] & [C/H] & [N/H] & [O/H] & [Ne/H] & [Mg/H]\\ \hline
2.0 & 0.0014 & ATM\_B & 0.5818 & 0.7334 & 11.5578 & 8.2546 & 8.4186 & 8.9395 & 8.2953 & 8.5866 \\
2.0 & 0.0014 & ATM\_C & 0.6251 & 0.7334 & 11.5578 & 8.2545 & 8.4188 & 8.9395 & 8.2953 & 8.5866 \\
2.0 & 0.0014 & ATM\_D & 0.6019 & 0.7334 & 11.5578 & 8.2545 & 8.4188 & 8.9395 & 8.2953 & 8.5866 \\
\multicolumn{11}{c}{…}                                             \\
8.0 & 0.014  & RES\_D & 0.5791 & 0.7079 & 11.5942 & 9.3008 & 9.4769 & 9.9329 & 9.3101 & 9.6015 \\
8.0 & 0.014  & RES\_E & 0.5821 & 0.7078 & 11.5944 & 9.3027 & 9.4753 & 9.9331 & 9.3102 & 9.6015 \\
\hline
\end{tabular}
\label{tab:abund}
\end{table*}

\subsection{Major Convective Regions of the Models and Their Extent}
\label{sec:extent}

In regions where convective transport of energy dominates, the matter is mixed almost instantly because of a very short thermal time scale of convection. Convective regions are thus chemically homogeneous. Typically, a discontinuity in the chemical profile is left at the border of the convective region. Throughout the evolution of an intermediate mass star, until the end of core helium burning, two major convective regions appear and alter the chemical profiles, the convective core on the MS and during core helium burning, and the convective envelope (CE) on the RGB (and later also during the AGB) -- see e.g., the Kippenhahn diagram in Fig.~\ref{fig:kip0}. The latter zone has a particularly important effect of bringing the nuclear-processed matter to the surface. We note that AGB evolution is particularly complex \citep[see e.g.][]{Herwig-2005}, requiring dedicated study; discussion below refers to RGB CE only.

\begin{figure*}
    \centering
    \includegraphics[width=\linewidth]{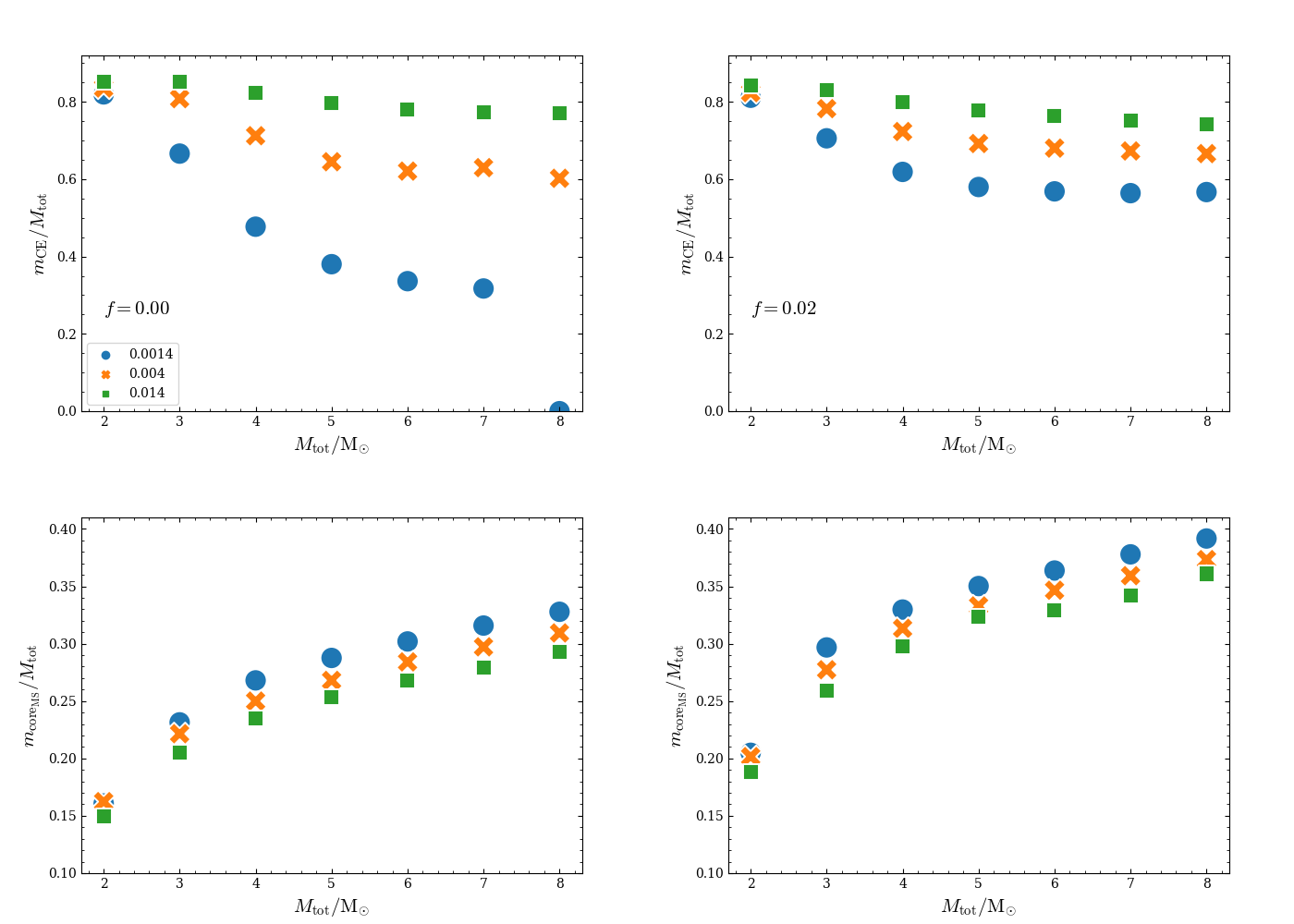}
    \caption{Maximum mass extent of the two largest convective regions: the envelope on the RGB (top panels) and the main sequence convective core (bottom panels), for the reference models. We plot $m_{\rm CE}/M_{\rm tot}$ and $m_{\rm core_{\rm MS}}/M_{\rm tot}$ vs. total stellar mass of a model, $M_{\rm tot}$, in the top and bottom panels, respectively. Results for different metallicities are plotted with different symbols/colors. On the left are the results for the canonical models, and on the right are the overshooting ones.}
    \label{fig:conv-extent}
\end{figure*}

We define the maximum relative depth of the CE as the ratio of the maximal mass of the CE
to the total mass of the star, $m_{\rm CE}/M_{\rm tot}$. We present it for the reference models in the top panels of Fig.~\ref{fig:conv-extent} (canonical models on the left panels, and overshooting models on the right), as a function of mass and metallicity. It can be seen that the lower the mass and the higher the metallicity, the deeper (more massive) the CE. Especially, for the lowest $Z$, and $M\geq4\MS$, the relative mass of the envelope becomes significantly smaller and disappears completely for the 8\MS\ model. On the tracks of these higher mass and low metallicity models, the RGB also starts to shorten and disappears for 8\MS. We would like to point out here that this particular model features thin convective shells after the MS, which we discussed in detail in \citetalias[]{Ziolkowska-2024}, and hence its structure might not be reliable. In fact, all the 8\MS\ models feature the thin convective shells, but the ones in $Z=0.0014$ are the most extended. Overshooting from the convective core on the MS increases the mass of the CE, especially for the two lower metallicity values (top right panel in Fig.~\ref{fig:conv-extent}).

Another factor that plays an important role is the maximum extent of the convective core during the MS phase. Note that during the main sequence evolution convective core retreats (see Kippenhahn diagram in Fig.~\ref{fig:kip0}) and its maximal mass extent is recorded early on the MS. It can be seen on the bottom panel of Fig.~\ref{fig:conv-extent} that the core grows with model mass and slightly shrinks with increasing metallicity. For a given mass, higher metallicity corresponds to increased opacity in the envelope, making the star cooler and less luminous. The energy production rate in the core is adjusted: the core is cooler and energy production rate drops; the net effect is a slightly smaller convective core. In the case of models calculated with overshooting from the main sequence hydrogen burning core, the extent of the zone presented in Fig.~\ref{fig:conv-extent} includes the convective core itself and the overshooting region above it. Overshooting increases the relative core mass by about 15 per cent.

\subsection{Investigating the Differences -- A Case Study}

In general, we can split our model sets into two groups. In the first one, the main reason for surface abundance disparities is the varying depth of the convective envelope. This includes model sets ATM, CONV, and RES. In the second, the initial chemical setup, namely relative ratios of elements in the reference solar mixture and reaction rates of \cag\ and \npg, are responsible for the disparities, and the model sets are MIX and NET. For models comprising INT and MLT sets, we do not observe any significant abundance differences compared to reference models.

To illustrate with an example of the $5\MS$, $Z=0.014$ model, significant abundance differences occur between the reference model and the models with modified nuclear network (NET\_B, NET\_D, NET\_E), different initial mixture of metals (MIX\_B, MIX\_C), and a different criterion for determining the convective boundary (CONV\_B). Here we focus on the latter, comparing a model in which the Ledoux criterion was used to set the convective core boundary, with a reference model in which the Schwarzshild criterion was used (CONV\_A; see Tab.~\ref{tab:labels}).

\begin{figure*}
    \centering
    \includegraphics[width=\linewidth]{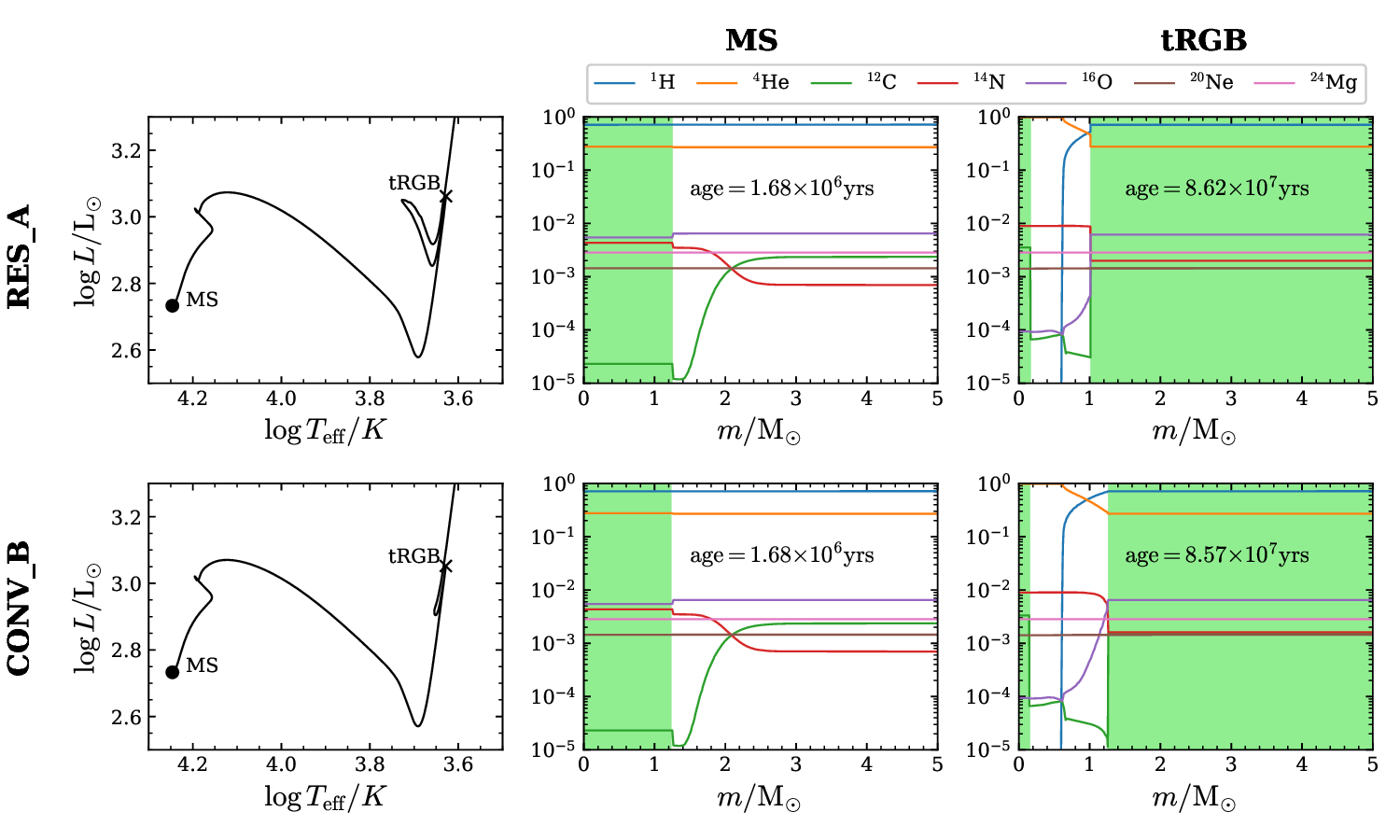}
    \caption{Comparison of the reference model (top) and a modified model with the Ledoux criterion (bottom). Left: HRDs with a point marking tRGB, the point of maximal extent of the CE. Middle: profiles with all the isotopic abundances on early MS, when the convective core is the most massive. Right: profiles on tRGB.}
    \label{fig:C-comp}
\end{figure*}

In Fig.~\ref{fig:C-comp} we compare HR diagrams of the two models (reference in the top, and modified in the bottom panels) as well as their profiles at early MS and tRGB -- the moments when the major convective zones are the most extended. The main difference between the two tracks is the blue loop shape, a feature that is very susceptible to subtle changes in the structure of a model.
The MS profiles look almost identical, and remain so until the first dredge-up occurs at tRGB, where the CE reaches to about $m=1\MS$ for the reference model and $m=1.2\MS$ for the Ledoux model. In both cases, it reaches the region where the abundances of C and O are lower and those of He and N are higher, effectively diluting the surface C and O and enhancing the surface He and N. The farther the CE reaches, the more carbon- and oxygen-poor, and nitrogen-rich matter is mixed to the surface. 
For each element, the differences between the reference and modified model amount to $\Delta$[C/H]=$-$0.02~dex, $\Delta$[N/H]=0.1~dex, $\Delta$[O/H]=$-$0.02~dex, $\Delta$[He/H]=0.01~dex, which is small, but, in principle, detectable. 

In the next section, we describe all the cases for which we note $|\Delta \rm [X/H]| \geq 0.01 dex$.

\subsection{Models Without Convective Core Overshooting}
We first discuss only the models without convective core overshooting.

\subsubsection{Convective Boundary Criterion, Schwarzschild vs. Ledoux}

Changing the convective boundary criterion from Schwarzschild to Ledoux, we notice differences in the surface abundances of several of our models. In the case of helium (differences of order, $\Delta$[He/H]$\lesssim$0.02), they are present for $Z=0.014$, $M\geq 3\MS$. 

The highest differences out of all the elements occur for carbon and nitrogen ($\Delta$[C/H]$\gtrsim$$-0.04$, $\Delta$[N/H]$\lesssim$$0.11$) for 3, 4\MS\ for all metallicities and 5, 6, 7, and 8\MS\ for $Z=0.014$. 

Oxygen abundances are affected in the following models: 3\MS, $Z=0.004, 0.014$ and 4, 5, 6, 7 and 8\MS, $Z=0.014$ (typically $\Delta$[O/H]$\gtrsim$$-0.02$).

In principle, the Ledoux and Schwarzschild criteria should result in the same convective boundaries when applied correctly on the convective side of the boundary. The boundary can be dislocated if the stability criteria are applied on the radiative side or when one interpolates for a sign change of $y=\nabla_{\rm rad}-\nabla_{\rm ad}$ when this function is discontinuous -- see \cite{Gabriel-2014} for a detailed discussion. In practice, proper determination of the convective boundaries is a very difficult and actual topic of research, see e.g., \citet{Paxton-2018}, \citet[][Sect.~3.3]{Ziolkowska-2024}. From any of the profiles that we show in this paper (e.g., in Fig.~\ref{fig:Tmix-profiles}), one can notice that the Ledoux gradient (blue continuous line) is far from smooth. It is calculated based on the spatial derivative of the distribution of all the isotopes. All the small distortions add up, resulting in the profile that often looks jagged with many sharp maxima in the radiative regions. In some cases, these maxima near the convective boundary result in splitting the convective region and a wrong determination of the boundary, and thus, a different depth of the CE. For this reason, we advocate for using the Schwarzschild criterion when studying surface abundances of stellar models, in particular at the lower boundary of the convective envelope.

\subsubsection{Numerical Setup, Refined Reference vs. Crude, Default}

In \citetalias[]{Ziolkowska-2024}, we performed a convergence study to determine a proper numerical resolution for our models. We varied temporal and spatial resolution as well as additional controls such as limits on change in central temperature and density, or distance between two consecutive points (time steps) in the evolution. The default \texttt{MESA} numerical setup is not optimized for science cases but rather for speed of calculations. 
Only for this crude numerical setup, we note two models with affected abundances, namely 3\MS, $Z=0.0014$ ($\Delta$[C/H]$\approx$$-0.01$, $\Delta$[N/H]$\approx$$0.01$) and 4\MS, $Z=0.0014$ ($\Delta$[N/H]$\approx$$0.01$). 

The different abundances are caused by a slightly different depth of the CE, which depends on the parameters controlling the numerical convergence of the model. We stress that for the other models in this set, we do not see any significant abundance differences. The numerical setup of our reference model that was chosen to ensure convergence of the evolutionary tracks, also ensures surface abundance convergence.

In the case of 8\MS, $Z=0.004$ all the modified resolution models produce noticeable abundance differences, some higher, some lower than the reference ($|\Delta$[C/H]|$\lesssim$ 0.03, $|\Delta$[N/H]$|\lesssim$ 0.04). However, this particular model is affected by the thin post-MS convective shells, described in \citetalias[]{Ziolkowska-2024}.

\subsubsection{Atmosphere, surface boundary interpolated from model atmosphere tables vs. integrated with different $T$-$\tau$ relations}

The surface boundary conditions, pressure, $P_{\rm s}$,  and temperature, $T_{\rm s} $, in the outermost cell of the model, can either be interpolated from model atmosphere tables, or obtained through direct integration of the hydrostatic balance equation, using one of the several available $T$-$\tau$ relations \citep{Paxton-2011}. In the reference model, we use model atmosphere tables that return $T_{\rm s} \equiv T_{\rm eff}$ and assume $\tau_{\rm s}=1$ and were computed using GN93 solar abundance mixture. In the varied models, the integration of the hydrostatic balance equation is carried out assuming different base of the atmosphere $\tau_{\rm b}=(0.3 \-- 2/3)$ and different $T$-$\tau$ relations, resulting in a slightly different surface boundary which then affects the underlying layers too.

Here we note small differences for 6 and 7\MS\ ($Z=0.004$; $\Delta$[C/H]$\approx$$-0.01$, $\Delta$[N/H]$\approx$$0.01$), and a slightly higher difference for the 8\MS, $Z=0.004$ ($\Delta$[C/H]$\approx$$-0.02$, $\Delta$[N/H]$\approx$$0.02$) for the model marked ATM\_C, with the $T$-$\tau$ relation by \citet{Krishna-Swamy-1966} and the lowest $\tau_b$ value. These are the same models that produced the highest differences in $\log T$ and $\log L$ in the ATM set on the evolutionary tracks in \citetalias[]{Ziolkowska-2024}. Again, a slightly different depth of the convective envelope is the direct cause of the difference.

\subsubsection{Heavy Element Mixture, A09 vs. GS98 and GN93}

By assumption, the models with different distributions of heavy elements have different initial mass fractions of H, He, and metals.
In this case, the differences do not reflect the modelling uncertainty; the reference solar metal mixture is a matter of choice and the abundance differences reflect the uncertainty of the solar composition.

\subsubsection{Nuclear Net and Reaction Rates} 
    
Another key factor that affects abundances is the network of isotopes and reactions that we choose to consider in the evolutionary modeling. As the network expands from the reference 25 isotopes to 49 in \texttt{mesa49.net} (`NET\_B'), the share of Ne and Mg abundances decreases by about 0.01--0.03~dex, regardless of mass or metallicity, to accommodate other elements in the bigger net (so the total $Z$ stays the same).

Another important aspect is the rate of two particular reactions, namely the reaction in the CNO cycle, \npg, and a helium-burning reaction, \cag. The default reaction rates in \texttt{MESA} (r-21.12.1) come from the NACRE compilation \citep{Angulo-1999}, but in the reference models we use newer, lower rates for these two reactions (see Tab.~\ref{tab:labels}). 

Using the \cag\ reaction rate by \citet{Angulo-1999} (NET\_C) at solar metallicity affects the core composition, lowering the $^{16}$O and raising $^{12}$C content in the center of the core. As an effect, the central C/O ratio can be lower by up to 27 per cent, as we discuss in detail in Sect.~\ref{sec:CO} and as mentioned in \citetalias[]{Ziolkowska-2024}.
    
The models calculated with the higher \npg\ rate from \citet{Angulo-1999} (`NET\_D', `NET\_E') are systematically different than those with the \citet{Cyburt-2010} rate (`NET\_A', `NET\_B', `NET\_C'). The slower CNO cycle makes the main sequence phase longer. More central $^{12}$C (by $\approx$65 per cent) and $^{16}$O (by $\approx$ 85 per cent) remains after the main sequence phase, regardless of mass or metallicity, while central $^{14}$N is the same for the different reaction rates.

Concerning the surface abundances, again the different depth of the convective envelope causes differences on the RGB and after. In the case of 6\MS, $Z=0.0014$, e.g., $m_{\rm CE}/M_{\rm tot}=0.34$ for the reference model and 0.42 for the modified model. Surface carbon and nitrogen are affected. The differences in surface carbon abundances occur in more models. Small ones ($\Delta$[C/H]$\approx$$-0.01$) occur for $Z=0.014$ (all $M$),  $Z=0.004$ for ($2, 3, 8$\MS) and the higher ones occur for $Z=0.0014$ and $4, 5, 6, 7$\MS\ ($\Delta$[C/H]$\lesssim$$0.05$, $\Delta$[N/H]$\gtrsim$$-0.11$).

\subsection{Models with Main-sequence Convective Core Overshooting}
\label{ssec:overshoot}

Overshooting from the hydrogen-burning convective core extends the main sequence lifetime and it makes the evolutionary tracks systematically more luminous. It affects the blue loops significantly. In the case of solar metallicity, it produces loops that are long enough to enter the instability strip ~\citepalias[see Sec.~3.2, Fig.~4 of][]{Ziolkowska-2024}.

\subsubsection{Comparison Between the Reference Canonical and Reference Overshooting Models}

First, we briefly discuss the differences between the surface abundances of the reference models with and without overshooting. For models with overshooting, the surface abundances diverge only after the dredge-up and the differences remain constant throughout the CHeB phase, similarly as before.

For the solar metallicity, the overshooting models have significantly higher helium ($\Delta$[He/H]$\approx-$0.02~dex), carbon ($\Delta$[C/H]$\approx-$0.01~dex) and nitrogen ($\Delta$[N/H]$\gtrsim-$0.07~dex) abundances, and lower oxygen abundance ($\Delta$[O/H]$\approx$0.01~dex, for $M\geq$ 3\MS). The above is also true for 2\MS\ model, for [He/H] and [N/H], but not [C/H] or [O/H].

In the lower metallicity models, $Z=0.004$, and overshooting case, helium abundance only differs for 2, 3\MS\ models, being higher than for the canonical models ($\Delta$[He/H]$\approx-$0.01). Carbon abundance is lower than the canonical value ($\Delta$[C/H]$\lesssim$0.06~dex, for $M\geq4\MS$, a difference growing with stellar mass). Nitrogen abundance is higher ($\Delta$[N/H]$\lesssim-$0.11, for all $M$, a difference growing with stellar mass). Oxygen abundance is lower ($\Delta$[O/H]$\approx$0.01~dex, for $M\geq$3\MS). 

For the lowest metallicity, $Z=0.0014$, helium abundance does not differ significantly, carbon is lower ($\Delta$[C/H]$\lesssim$0.2 for models with $M\geq$3\MS, and the difference grows with model mass). Nitrogen is higher ($\Delta$[N/H]$\lesssim$0.37, a difference growing with stellar mass).

These differences reflect the different extent of the main convective zones of the models, both the MS core and the CE. As shown in Fig.~\ref{fig:conv-extent}, overshooting models have extended MS cores, but also CE, especially for the lowest metallicity value (as discussed in Sect.~\ref{sec:extent}), here the
differences are the largest, as compared to no overshooting
models, the envelope is on average by a factor of 0.17 deeper (an average of the difference between the models of 3--7 \MS).

\subsubsection{Differences Between the Reference Overshooting Models and Modified Overshooting Models}

With a few exceptions, changing the input modeling setup of the overshooting models has a qualitatively similar effect on surface abundances as in the case of the canonical models. Below we list all the overshooting models for which $|\Delta$[X/H]|$\geq$0.01~dex.

\begin{enumerate}
 \item In the CONV set,
several models calculated with the Ledoux criterion have a different surface helium abundance than the reference model. This is the case in 3, 4, 5\MS\ and $Z=0.014$ models (similar to the canonical models), but no difference is observed for [He/H] for higher masses at solar metallicity. [He/H] also differs for 2, 3\MS\ and $Z=0.004$ ($\Delta$[He/H]$\lesssim$0.03~dex). 

Carbon is affected for 3\MS\ for all the metallicity values, and for $M\geq 4 \MS$ for $Z=0.004$ ($\Delta$[C/H]$\approx$$-0.01$~dex).

Nitrogen abundances are affected across all stellar masses. In 2-3 \MS\ models, this effect is present at all metallicities, whereas for $M \geq 4 \MS$ it appears only at $Z = 0.004$ and $Z = 0.014$, with $\Delta$[N/H] $\approx$ 0.13 dex. The strongest differences are found at lower masses and Z = 0.004.

Oxygen is affected for 3\MS\ and $Z=0.004$ and $Z=0.014$ as well as for $M=4,7,8\MS$ and $Z=0.004$ ($\Delta$[O/H]$\gtrsim$$-0.02$~dex). 

Additionaly, for one model, $3\MS$, $Z=0.014$, neon and magnesium are affected ($\Delta$[Ne/H]$\approx\Delta$[Mg/H]$\approx$0.01).

\item 
In the case of the RES set, the only models affected are $M=3, 4, 5\MS$ and $Z=0.004$ and $5\MS$ and $Z=0.014$, where $\Delta$[N/H]$\approx$0.01~dex in the model calculated with the crude default MESA numerical setup (RES\_E). 

\item 
In the ATM set, nitrogen abundance is the most diverging from the reference, typically by $\Delta$[N/H]$\approx$0.01~dex, mostly for $Z=0.004$, for masses $M=2,4,5,6,7,8 \MS$, but also for $M=6,7\MS$ and $Z=0.0014$. The differences appear for models with the same $T$-$\tau$ relations as in the canonical models (ATM\_C, ATM\_D, and in one case ATM\_E for $5\MS$, $Z=0.004$). 

Differences in carbon ($\Delta$[C/H]$\approx$$-0.01$~dex) occur for models $M=5,6\MS$ and $Z=0.0014$ for Krishna-Swamy (ATM\_C) $T$-$\tau$ relation. 

\item 
For the NET set, carbon and nitrogen are affected. Carbon is affected in more models. For carbon those are $M=2$\MS\ and all the metallicity values; the models with $M>3\MS$ and $Z=0.004$ and $Z=0.014$ ($\Delta$[C/H]$\approx$$-$0.01~dex). Nitrogen abundance in the NET\_D and NET\_E models, as compared to the reference models, is either higher (for $Z=0.004$ and $M\geq\MS$) or lower (for $Z=0.014$ and $M=2, 3, 4, 5, 7\MS$) by $|\Delta$[N/H]$|\approx$0.01~dex.  

\end{enumerate}

Overall, the surface abundance differences between the reference and modified models with overshooting follow the same trends as in the case of the canonical models. The underlying reason behind the differences is also the same. The masses of the convective envelopes of the overshooting models are higher, especially for the $Z=0.0014$ models, which is reflected in the evolutionary tracks, which have longer RGB.

\subsection{Central C/O Ratio at the End of Core-Helium Burning}
\label{sec:CO}

The ratio of central $^{12}$C/$^{16}$O in intermediate-mass stars \citep[$M\lesssim 9\MS$ according to][]{Umeda-1999} is one of the most important factors affecting white dwarf evolution, and later the brightness and diversity of Type Ia Supernovae \citep[e.g.,][]{Xiao-2020, Umeda-1999, Hoflich-1998}. 

The main factor controlling the C/O ratio is the relative rates of two reactions, firstly the triple-$\alpha$ reaction, fusing three helium nuclei into $^{12}$C, and then, after enough carbon is produced, the alpha capture by carbon, \cag.

\begin{figure*}
    \centering
    \includegraphics[width=\textwidth]{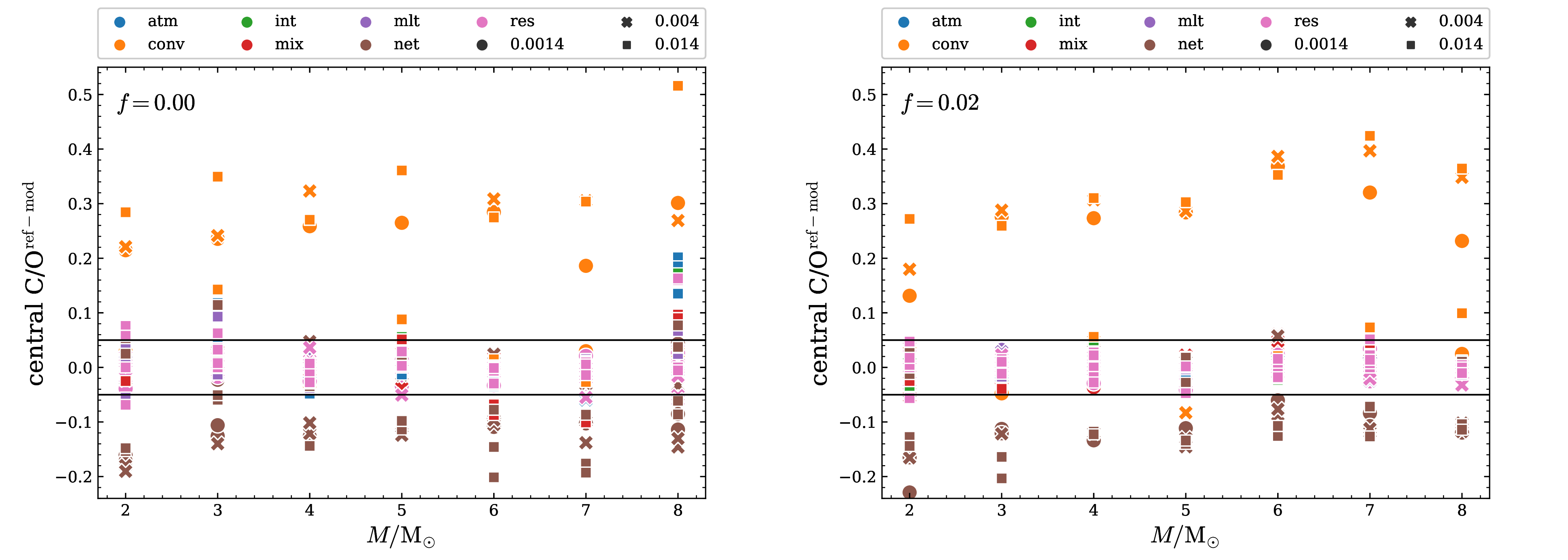}
    \caption{Difference of central C/O ratio at eCHeB between the reference and modified models in our grid. On the left are canonical models, and on the right are the overshooting ones. The horizontal line marks central C/O$^{\rm ref-mod}$=0.05. Colors correspond to different sets and symbols to different metallicities, according to the key above the panels.}
    \label{fig:CO_comp}
\end{figure*}

The difference of the C/O ratio at the eCHeB between the reference and modified models is presented in Fig.~\ref{fig:CO_comp}, with dependence on mass (x-axis), metallicity (symbol), and the model set (color). We discuss the cases where the C/O ratio of a given model differs by $|\Delta$C/O$|\geq$0.05 from the reference model value.

The CONV set has the strongest impact on the central C/O ratio, in particular CONV\_C, marking models calculated with a simple sign change algorithm for determining convective boundaries. Using this algorithm effectively lowers the C/O ratio by, on average, $\Delta{\rm C/O}\approx 0.3$, or up to 0.5 for the 8\MS, $Z=0.014$ (a model whose structure might not be reliable due to thin convective shells present). The second most important model set is NET. In particular, choosing the \citet{Angulo-1999} \cag\, reaction rate over \citet{Kunz-2002} increases the C/O ratio by $\Delta {\rm C/O}\approx 0.1-0.2$, depending on mass and metallicity. Changing the rate of \npg\ reaction affects the C/O ratio for 6 and 7\MS\ but only slightly, and only for the canonical models. 

Modifying atmospheres, MLT formalism, and numerical resolution does have a noticeable effect on central C/O ratio in single cases, but it is small ($\Delta{\rm C/O}\lesssim 0.1$) with the exception of the 8\MS\ and $Z=0.014$ model for which $\Delta{\rm C/O}\lesssim$0.2.

In general (except some models), including overshooting (right panel in Fig.~\ref{fig:CO_comp}) slightly decreases the overall average difference between the C/O in the reference and modified models.

\begin{figure*}
    \centering
    \includegraphics[width=\linewidth]{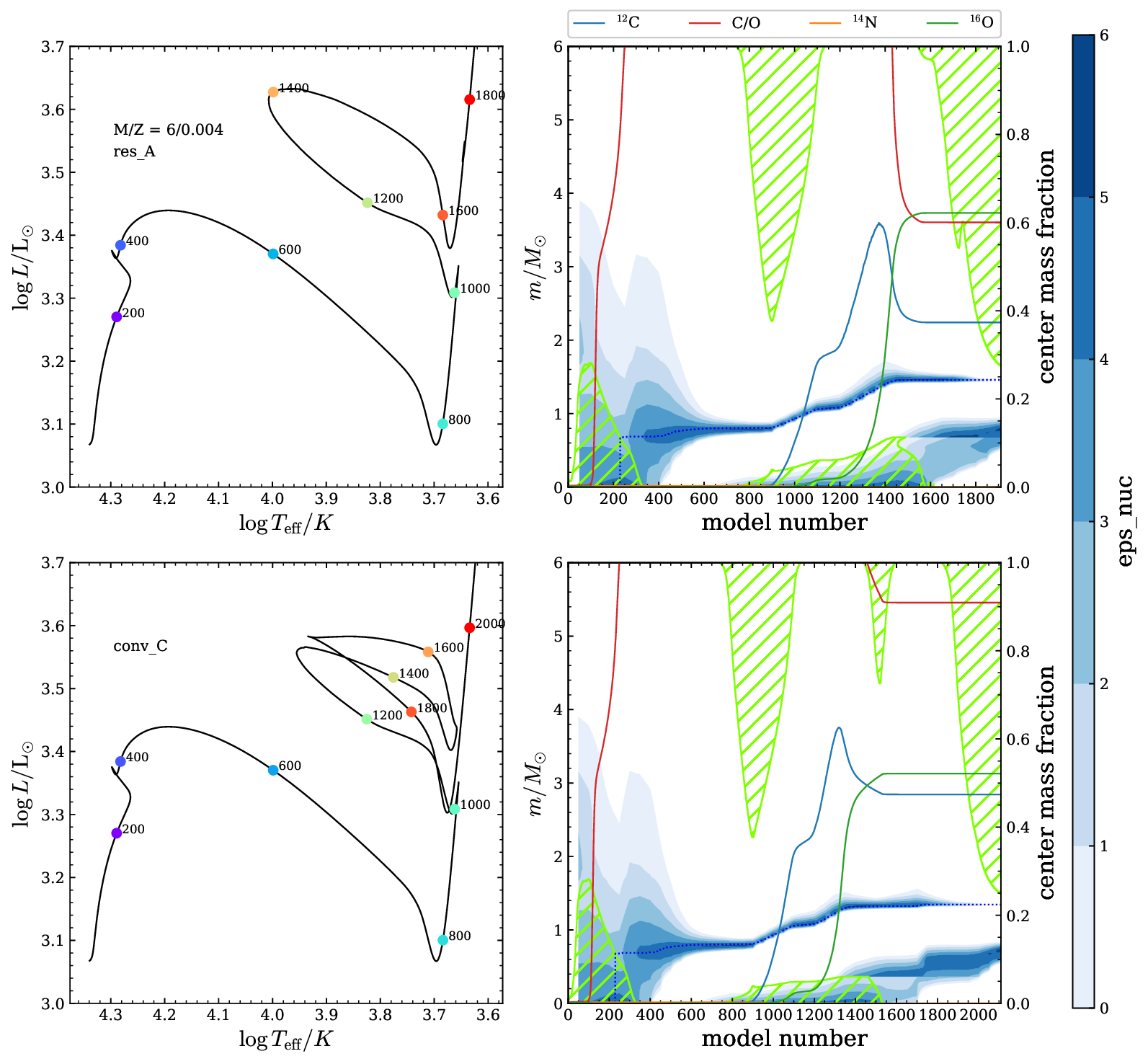}
    \caption{Comparison of the reference model with predictive mixing scheme for determining the convective boundaries (top) and modified model with sign change algorithm, CONV\_C (bottom). On the right are Kippenhahn diagrams, with model numbers on the x-axis, and on the left HRDs with markers for every 200th model.}
    \label{fig:COkipp} 
\end{figure*}

The highest differences for central C/O values at eCHeB in our whole grid appear for models marked CONV\_C which are the ones using a simple sign change algorithm  \citep[see e.g., Sect.~2][]{Paxton-2018} for locating the boundary of the convective core. In Fig.~\ref{fig:COkipp} we compare the two models on HR and Kippenhahn diagrams. Using the sign change algorithm results in a smaller helium-burning core, encompassing less mass. The difference in the core size causes a slightly different composition in these two models that starts to be apparent from the model number of about 1100, where a small plateau appears in the carbon profile in Fig.~\ref{fig:COkipp}. At the end of core-helium burning, the core is split, and the triple-$\alpha$ reactions occur mainly in a thick shell above the core. From that point on, the central $^{12}$C and $^{16}$O abundances stay constant. For the CONV\_C model, the convective core vanishes sooner, and C/O stays at a higher level than in the reference model. Additionally, the tracks are also affected in the CONV\_C model, a secondary blue loop develops, accompanied by a convective envelope seen in the Kippenhahn diagram (bottom right panel on Fig.~\ref{fig:COkipp}. We note that, as explained by \cite{Gabriel-2014} and demonstrated by \cite{Paxton-2018}, the use of sign change algorithm at the boundary with composition discontinuity leads to incorrect determination of the boundary location. As evident from Fig.~\ref{fig:COkipp}, it also has significant consequences for blue loop evolution. The correct treatment of convective boundaries is essential for this and other stages of evolution.

\section{Discussion and Conclusions}
\label{sec:discussion}

We used \texttt{MESA} version r-21.12.1 to calculate the evolutionary models of $M=2\--8\MS$ and $Z=(0.0014,0.004,0.014)$ with 22 modified physical and numerical setups, from ZAMS until core-helium exhaustion, with and without main-sequence convective core overshooting. We provide the surface abundances and central mass fractions of basic elements, $^1$H, $^4$He, $^{12}$C, $^{14}$N, $^{16}$O, $^{20}$Ne, $^{24}$Mg, as well as central $\rm ^{12}C/^{16}O$. We list and describe the cases where the difference in the surface abundances between the reference and modified models is greater than $\sim0.01$~dex.

\begin{figure*}
    \centering
    \includegraphics[width=\textwidth]{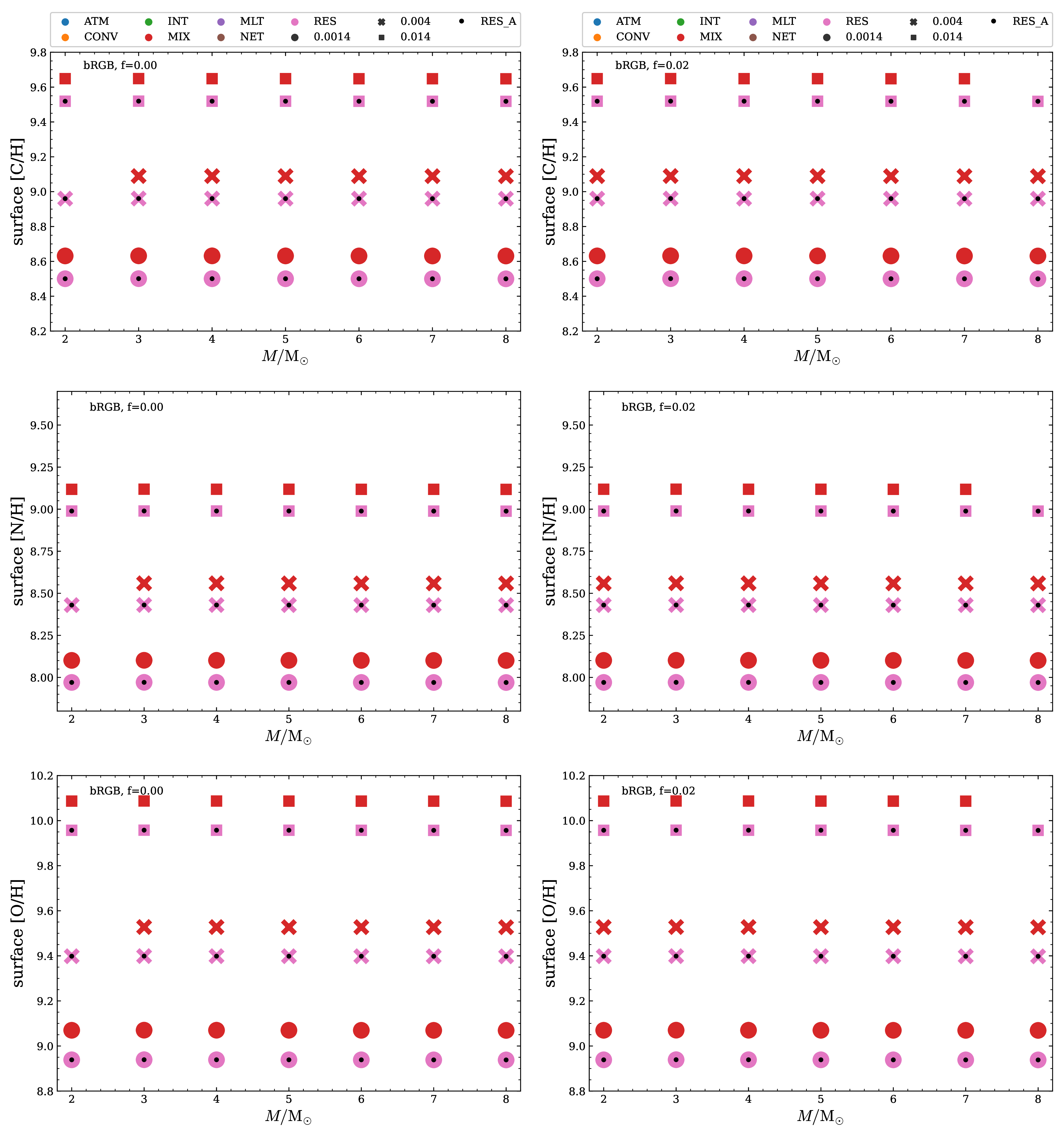}
    \caption{Abundances of C, N and O before the first dredge-up, at the base of the red giant branch, as a function of mass (x-axis) and metallicity (symbols; squares for $Z=0.014$, crosses for $Z=0.004$ and circles for $Z=0.0014$) and model set (color) for canonical (left panels) vs. convective core overshooting (right) models. The reference model values are marked with a black dot and most often overlap with the RES\_E (pink) points. For each set, we chose one model that stands out the most. Still, most of the points for given $M/Z$ are obscured by each other, not differing significantly.}
    \label{fig:CNO-bRGB}
\end{figure*}

\begin{figure*}
    \centering
    \includegraphics[width=\textwidth]{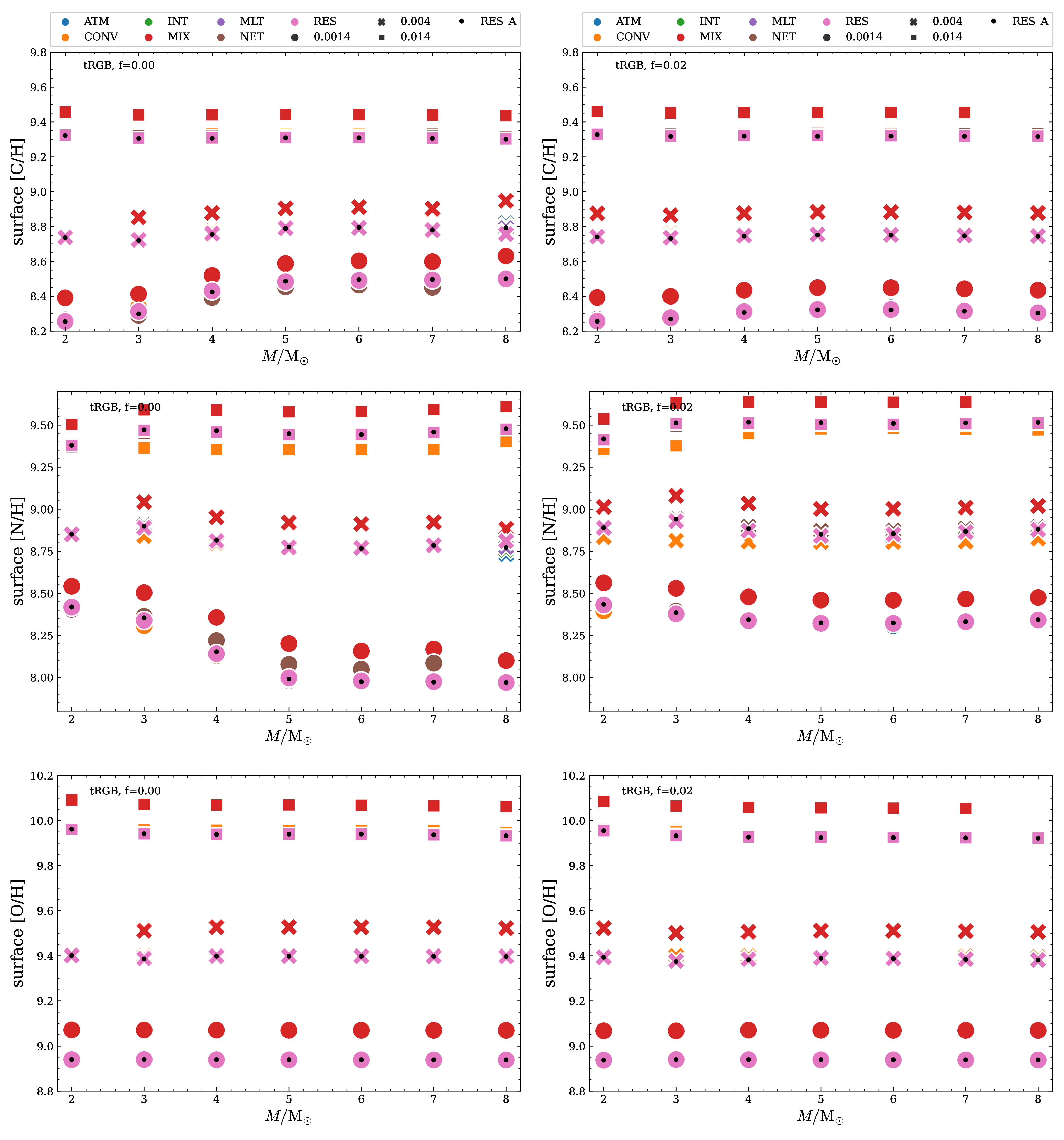}
    \caption{The same as Fig.~\ref{fig:CNO-bRGB} but for tRGB.
    }
    \label{fig:CNO-tRGB}
\end{figure*}

We collect the results in Figs~\ref{fig:CNO-bRGB} and \ref{fig:CNO-tRGB} which show
surface abundances of $^{12}$C, $^{14}$N, and $^{16}$O at bRGB (before the first dredge-up) and tRGB (after the dredge-up) as a function of mass (x-axis), metallicity (symbol), model set (color), and overshooting (left vs. right panels). Abundances in the reference models are marked with a black dot. Then, from each model set, the most divergent model is selected and plotted. Most of the points are obscured because the differences in abundances are small.

Before the first dredge-up (Fig.~\ref{fig:CNO-bRGB}), as discussed above, the surface abundances are marginally different in the considered model sets. The only relevant differences between any two models of a given $M/Z$ come from changing the solar mixture of elements (e.g., MIX\_C, which stands for the GN93 solar mixture model). No other aspect of modelling that we varied did affect the surface abundances significantly before the dredge-up. Hence, most of the points are hidden below the pink symbols corresponding to the RES\_E model (default numerical resolution controls), which is the most divergent in the RES set (still typically below 0.01\,dex).

After the dredge-up, the effect of different modelling assumptions becomes relevant for surface abundances as can be seen in Fig.~\ref{fig:CNO-tRGB}. The highest differences in surface abundances, reaching 0.1\,dex, occur for [N/H] and [C/H] after modifying the reference solar metal mixture to GN93 (MIX\_C), changing reaction rates of \cag\ (NET\_C, NET\_E) and convective boundary criterion to Ledoux (CONV\_B). We see a dependence on model mass both for the abundances of C and N and for the abundance differences with respect to the reference model. The mass dependence is especially seen for the models for which the envelope of the reference model is very small --- the models without overshooting and $Z=0.0014$ (see \ref{sec:extent}). The reference non-overshooting models with $M=2,3\MS$ and all the reference overshooting models have [C/H]$\approx$8.2 and [N/H]$\approx$8.5, while the $M\geq4\MS$ standard models have [C/H]$\approx$8.4--8.5 and [N/H]$\approx$8.0. The abundance differences grow with model mass - we see a bigger spread with growing model mass, especially for the models listed above. 

In the remaining sets, RES and ATM, the highest differences correlate with the differences in the evolutionary tracks. Often, we note the significant differences for 8\MS\ model. The elements that are affected the most and most often are [N/H], [C/H], more rarely [He/H], rarely [O/H], and almost never [Ne/H] or [Mg/H]. When using the crude default numerical resolution controls (RES\_E), the differences appear for the lowest metallicity value and $M=3,4\MS$. When using different $T$-$\tau$ relations, they appear for $Z=0.004$ and $M=6,7,8\MS$. For models with overshooting, the cases with $\Delta$[X/H]$\geq$0.01 within RES and ATM sets are more numerous, but the differences are similar to those in the canonical models.

In the sets MLT, where we changed the MLT formalism and INT, where we varied the interpolation scheme of opacity tables between cubic and linear in $X$ and $Z$, we do not record any significant differences in surface abundances as compared to a reference model. Both the evolutionary tracks and surface composition are robust against these numerical choices. 

\begin{figure*}
    \centering
    \includegraphics[width=\textwidth]{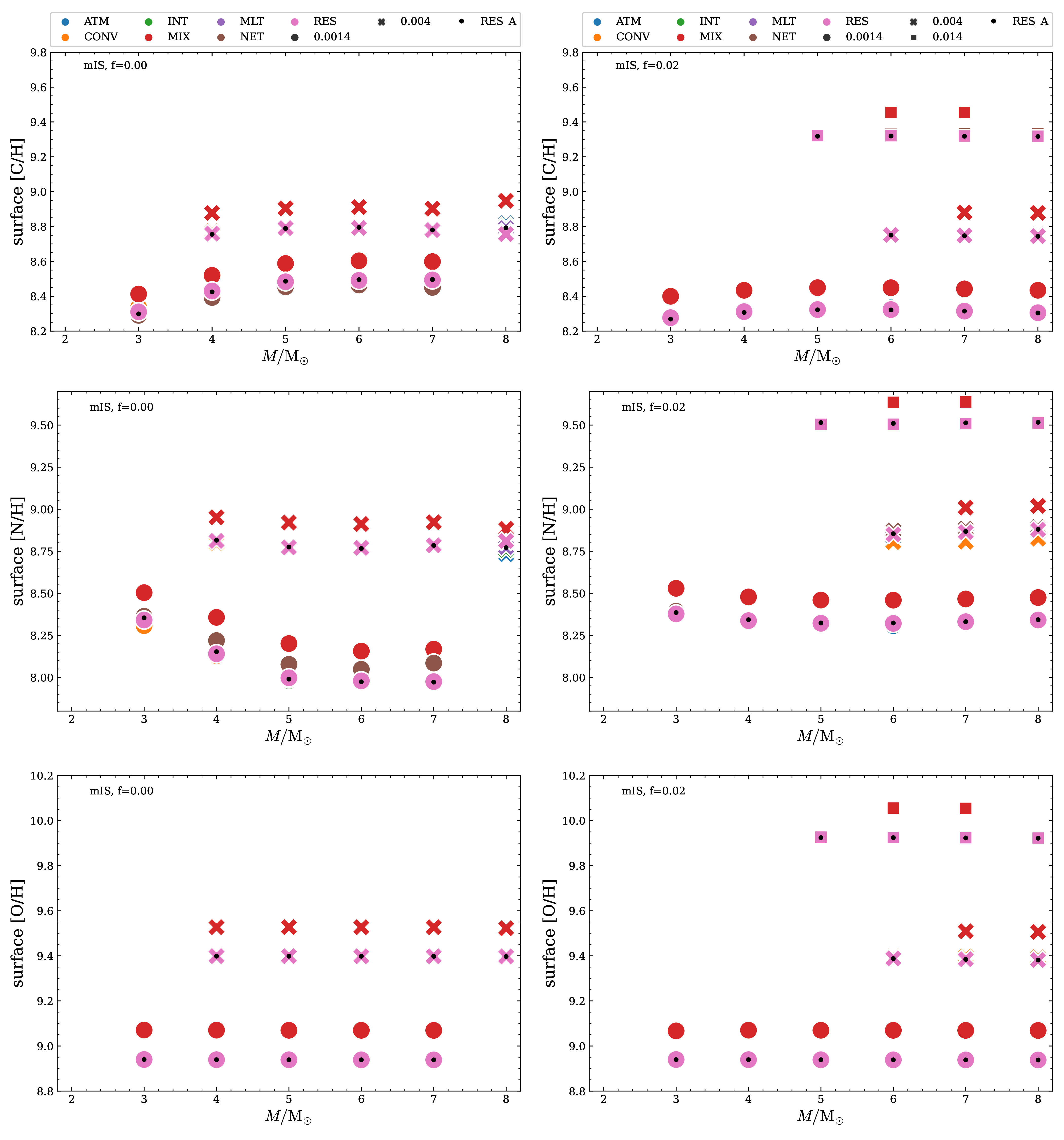}
    \caption{The same as Fig.~\ref{fig:CNO-bRGB} but for mIS.
    }
    \label{fig:CNO-mIS}
\end{figure*}

In the context of classical Cepheids, we include Fig.~\ref{fig:CNO-mIS} with differences in $^{12}$C, $^{14}$N, and $^{16}$O at mIS, the midpoint of the classical instability strip. The models that do not enter the IS (all 2\MS\ models; all solar metallicity models, $8\MS$/$Z=0.0014$ and $3\MS$/$Z=0.004$ models with no overshooting; $M<6\MS$/$Z=0.004$ and $M<5\MS$/$Z=0.014$ models with overshooting), are absent from the figure. Except for those missing models, the trends remain very similar as at tRGB (Fig.~\ref{fig:CNO-tRGB}).


Overall, the differences of surface chemical abundances are small. Their values are not very sensitive to the choice of parameters and are a robust outcome of the evolutionary calculations.
Still, some differences are noticeable. In the case where we varied the reaction rates, or the solar mixture of heavy elements, the discrepancy in abundances are obvious and expected. Then, the largest differences were encountered in the CONV set, in which prescriptions to define convective boundaries were explored. The differences reach up to 0.11\,dex ([N/H] for $M=4\MS$, $Z=0.014$). They reflect an important and actively investigated problem of proper determination of convective boundaries. In case of other sets of models, like varied atmospheres, or numerical resolution controls, the spread in abundances is very small (typically below 0.01\,dex; in the most extreme case 0.04\,dex, and the major {\it{physical}} factor behind the differences between models is the different depth of the convective envelope zone on the RGB.

Our models do not include mass loss and rotation and were computed with a rather coarse step in mass and metallicity, $\Delta M =1\MS$ and $\Delta{\rm [Fe/H]}=0.5$. As such, they are not intended for detailed comparison with observations or to constrain some model parameters. The main purpose of this study was to quantify to what extent the surface abundances are sensitive to the choices regarding physical and numerical formulation. In nearly all published studies, the factors that we explored are simply fixed and the effects of their changes are not discussed. This is rational, as other factors, such as overshooting at various convective zones, rotation and associated mixing processes, and mass loss more strongly impact the evolutionary tracks. These phenomena should be included when comparing with observations. However the knowledge on uncertainty arising due to these {\it secondary choices} we have discussed is important to make reliable comparisons and draw meaningful conclusions. With the tracks we published online, one may also track the abundances through the whole evolution, till end of core helium burning, not only at specific benchmark points.

We also investigated the central C/O ratio at the end of core helium burning, which is important for later stages of evolution, e.g., in white dwarfs. The C/O ratio depends on two reactions simultaneously happening during helium burning, the triple alpha reaction and alpha capture reaction, \cag. Here, it is important to set the boundary of the convective core correctly, as using an inappropriate procedure may affect the size of the core and duration of central helium burning and, in turn, change the central C/O ratio by on average 40 per cent (up to 120 per cent). Another important aspect is the rate of the \cag\ reaction. Using the default \citet{Angulo-1999} rate vs. the newer, lower \citet{Kunz-2002} rate increases the C/O ratio by 0.1--0.2, both for overshooting and standard models. Other factors give rise to an uncertainty of $\Delta$C/O$\leq$0.05 or about 10 per cent.

It is evident that central C/O ratio is sensitive to the physical and numerical formulation of the model. Constraining the underlying physics, like nuclear reaction rates or schemes to determine convective boundaries through observations, in particular observations of only surface abundances, would be challenging, if possible at all. Mixing processes are complex and their effects manifest through many observables, which should be taken into account. Modeling components of eclipsing binary systems, for which high-resolution spectroscopic observations are available is certainly a promising avenue. The most promising approach is through asteroseismic inferences, that give a direct insight into stellar interior \citep[for a review, see e.g.,][]{Kurtz-2022}, however these are only possible for multiperiodic pulsators. One should also keep in mind that mixing processes are intrinsically 3D and further development of the modeling tools is necessary as well.

Finally, we would like to point out some inconsistencies in stellar evolution modeling related to abundances and pass on a few recommendations.

A well-known inconsistency is related to the computation of opacity. Opacity tables over which interpolation is computed assume a fixed chemical composition of metals. Other than using type II opacities that take into account enhanced C and O abundances at later evolutionary stages, the changing composition of elements contributing to metallicity, $Z$, is not taken into account. A similar problem is related to tables of atmosphere properties that are used to model the surface layers. Computations that are used to construct these tables assume a fixed metal content that may be different than the one used in opacity calculations. This is the case for a reference model set which uses atmosphere tables computed assuming GN93 solar abundance mixture \citep{Hauschildt-1999a, Hauschildt-1999b}, while to set initial chemical composition, and as a reference for opacity calculations, we use A09. While related uncertainties are usually assumed to be small, their explicit assessment is difficult due to a scarce set of atmosphere tables published and included in \texttt{MESA}.

We also note that the chemical composition of a particular star may differ from the scaled solar composition -- reflecting the environment in which the star formed and its individual evolution along with associated mixing processes.  In our calculations, as is common practice for model surveys, we simply assumed a scaled solar chemical composition.

Considering MESA modelling, we recommend careful inspection of Kippenhahn diagrams and abundance profiles, in particular at surface layers, as a problem of thin radiative shells at the surface, requiring to set additional controls (\texttt{T\_mix\_limit}), may be present. Depending on the problem investigated, the start of calculations at the pre-main sequence may also be necessary.

\section*{Acknowledgements}
This research is supported by the National Science Center, Poland, Sonata BIS project 2018/30/E/ST9/00598. OZ was supported by the National Science Center OPUS grant 2023/49/B/ST9/01671. This research was supported in part by grant NSF PHY-1748958 to the Kavli Institute for Theoretical Physics (KITP). A.T. is a Research Associate at the Belgian Scientific Research Fund (F.R.S.-F.N.R.S.).

\software{PyMesaReader \citep{Wolf-2017}, Mkipp \citep{Marchant-2019, Marchant-2020}, \texttt{MESA} 21.12.1 \citep{Paxton-2021}, \texttt{MESA} SDK  \citep{Townsend-2022}, Numpy \citep{Harris-2020}, Matplotlib \citep{Hunter-2007}, Pandas \citep{Reback-2022,McKinney-2010}}.

\appendix 

\section{Inlist for an exemplary reference model}
\label{appendix:inlist}

\begin{lstlisting}[language=fortran]
&star_job
! abundances
    initial_zfracs = 6 ! A09
    
! nuclear net  
    change_net = .true.
    new_net_name='pp_and_cno_extras.net'  
        
    set_rate_c12ag = 'Kunz'
    set_rate_n14pg = 'jina reaclib'
    
! chemical composition for ZAMS model
    relax_y = .true.
    new_y = 0.2695
    relax_z = .true.
    new_z = 0.014
/

&kap
    kap_file_prefix = 'a09'
    kap_lowT_prefix = 'lowT_fa05_a09p'
    kap_CO_prefix   = 'a09_co'

    use_Zbase_for_Type1 = .true.
    use_Type2_opacities = .true.
    Zbase = 0.014

    cubic_interpolation_in_X = .true.
    cubic_interpolation_in_Z = .true.
/

&controls
    initial_mass = 5

    MLT_option = 'Henyey'
    mixing_length_alpha = 1.77

! convective boundaries
    use_Ledoux_criterion = .false.

    recalc_mix_info_after_evolve = .true.

    predictive_mix(1) = .true.
    predictive_zone_type(1) = 'any'  
    predictive_zone_loc(1) = 'core'   
    predictive_bdy_loc(1) = 'any'     

    predictive_superad_thresh(1) = 0.005d0
    predictive_avoid_reversal(1) = 'he4'

    T_mixing_limit = 1d6
    
! semiconvection
    alpha_semiconvection = 0

! atmosphere
    atm_option = 'table'
    atm_table = 'photosphere'
    atm_off_table_option = 'T_tau'

! diffusion
    do_element_diffusion = .false.

! convergence parameters
    mesh_delta_coeff = 0.5d0  
    time_delta_coeff = 0.5d0
    max_years_for_timestep = 1d6

    varcontrol_target = 1d-4
    max_allowed_nz = 32000

    ! limit on magnitude of relative change at surface
    delta_HR_limit = 0.005d0   
    delta_lgTeff_limit = 0.005d0   
    delta_lgL_limit    = 0.01      

    ! limit on magnitude of relative change at center
    delta_lgT_cntr_limit   = 0.005d0  
    delta_lgRho_cntr_limit = 0.025d0   

    ! when to stop
    max_age = 15d9

    warn_when_large_rel_run_E_err = 99d0
    calculate_Brunt_N2 = .true.
/
\end{lstlisting}

\bibliography{main} 
\bibliographystyle{aasjournal}

\end{document}